\documentclass[conference]{IEEEtran}
\IEEEoverridecommandlockouts
\usepackage{cite}
\usepackage{amsmath,amssymb,amsfonts}
\usepackage{algorithmic}
\usepackage{graphicx}
\usepackage{textcomp}
\usepackage{xcolor}
\usepackage{hyperref}
\usepackage{booktabs, tabularx}
\usepackage{subcaption}
\usepackage{multirow}
\usepackage{balance}
\usepackage[braket,qm]{qcircuit}
\def\BibTeX{{\rm B\kern-.05em{\sc i\kern-.025em b}\kern-.08em
    T\kern-.1667em\lower.7ex\hbox{E}\kern-.125emX}}
\begin{document}

\bstctlcite{IEEEexample:BSTcontrol}


\title{Controlled Steering-Based State Preparation for Adversarial-Robust Quantum Machine Learning}

\author{\IEEEauthorblockN{Sahan Sanjaya, Hari Krishna Parvatham, Emma Andrews, and Prabhat Mishra}
\IEEEauthorblockA{
\textit{University of Florida, 
Gainesville, FL, USA}}
}

\maketitle

\begin{abstract}

Quantum machine learning (QML) provides a promising framework for leveraging quantum-mechanical effects in learning tasks. However, its vulnerability to adversarial perturbations remains a major challenge for practical deployment. In QML systems, small perturbations applied to classical inputs can propagate through the quantum encoding stage and distort the resulting quantum state, thereby degrading model performance. In this work, we propose a defense mechanism that replaces the conventional quantum encoding stage of a QML model with passive steering-based controlled state preparation, which guides the encoded state toward a controlled intermediate state. By tuning the steering strength and the number of steering iterations, the proposed method suppresses the influence of adversarial perturbations while maintaining high clean accuracy and improving adversarial accuracy. Experimental results demonstrate that the passive steering-based defense consistently improves adversarial accuracy across different QML models and datasets under gradient-based adversarial attacks, achieving adversarial accuracy improvements of up to 40.19\%.



\end{abstract}


\section{Introduction}

Quantum machine learning (QML) has emerged as a promising paradigm for leveraging quantum-mechanical principles in data processing and pattern recognition tasks. QML models, such as quanvolutional neural networks (QNNs)~\cite{henderson2020quanvolutional}, quantum convolutional neural networks (QCNNs)~\cite{cong2019quantum}, and variational quantum classifiers (VQCs)~\cite{schuld2015introduction}, have shown potential advantages in expressive capability and compact model representation. A fundamental component of QML models that operate on classical data is the quantum encoding stage, where classical inputs are mapped to quantum states through angle encoding, amplitude encoding, or related state-preparation techniques~\cite{khan2024beyond}. Since this stage defines the initial quantum representation supplied to the subsequent learning model, its design has a direct impact on the overall inference process.

\begin{figure}[t]
    \centering
    \includegraphics[width=\linewidth]{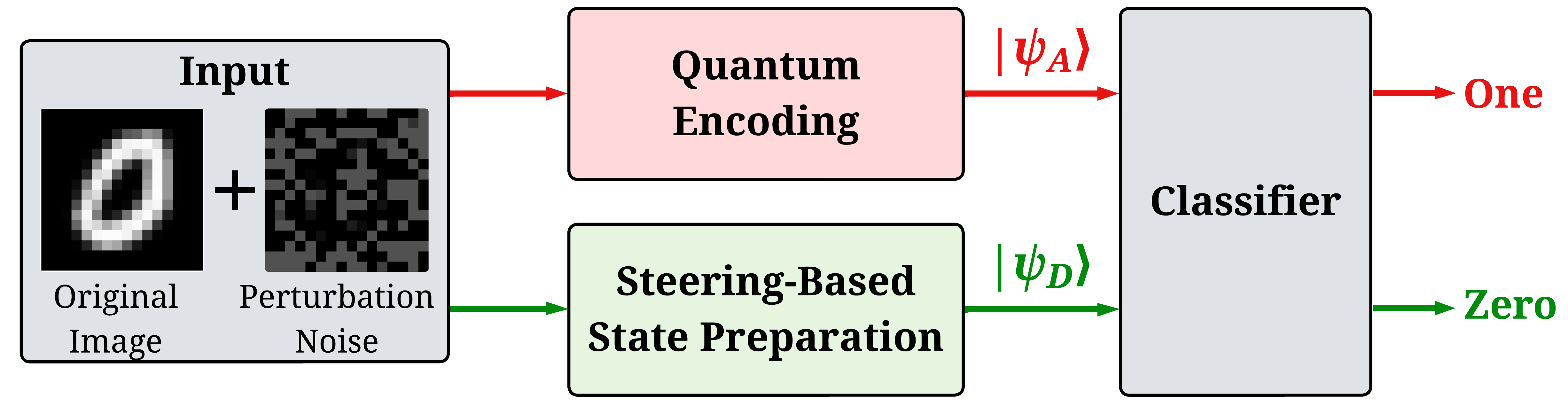}
    \caption{Controlled measurement-induced passive steering-based state preparation (green path) replaces the standard quantum encoding circuit (red path) as a defense against adversarial attacks. Here, \(|\psi_A\rangle\) denotes the adversarially affected encoded state produced by the standard encoding circuit, while \(|\psi_D\rangle\) denotes the controlled intermediate state generated by passive steering, which serves as both the defense mechanism and the encoding stage.}
    \label{fig:intro_method}
\end{figure}

Despite this promise, QML models remain vulnerable to adversarial attacks, similar to their classical counterparts~\cite{liao2021robust, anil2024generating, akter2024quantum}. In such attacks, carefully crafted input perturbations, although small in magnitude, can cause significant degradation in inference performance. As QML systems move closer to practical deployment, improving their resilience against adversarial attacks has become an important research challenge. As illustrated in Figure~\ref{fig:intro_method}, the red arrows depict the conventional adversarial attack path on a QML model that incorrectly classifies a `0' as `1' due to the noise at the input. Recent studies have explored several defense strategies against adversarial attacks in QML, including adversarial training~\cite{lu2020quantum,ren2022experimental}, regularization-based methods~\cite{wendlinger2024comparative,berberich2024training,li2026dual}, defensive noise injection~\cite{huang2023enhancing,du2021quantum,gong2024enhancing}, robust quantum circuit design~\cite{kananian2025partitioned,kananian2025adversarial,el2024robqunns}, and input preprocessing~\cite{khatun2025classical}. However, there are only a limited number of input preprocessing-based defenses that can be integrated into trained QML models without requiring retraining or model-specific architectural modifications.

To address this limitation, this paper proposes a defense mechanism based on a controlled measurement-induced steering-based state preparation framework that modifies the encoded quantum state before it is processed by the inference model. As shown in Figure~\ref{fig:intro_method}, the proposed approach exploits passive steering dynamics to guide the system towards a controlled intermediate state along the trajectory to the desired encoded state, where the behavior is governed by parameters such as the steering strength and the number of steering iterations. The key idea is to stop the steering process before the exact encoded target state is reached, so that the encoding stage itself acts as a robustness-oriented filter. This mechanism enables suppression of adversarial perturbations at the state-preparation stage while preserving clean-input performance to a large extent.

The proposed method is evaluated on representative QML models under gradient-based adversarial attacks, including the fast gradient sign method (FGSM)~\cite{goodfellow2015explaining} and projected gradient descent (PGD)~\cite{madry2019deep}, across standard image-classification datasets. Both clean and adversarial settings are considered in order to examine the trade-off between clean accuracy and adversarial robustness. Experimental results show that controlled passive steering can significantly improve adversarial accuracy with minor impact in clean accuracy when the steering parameters are properly selected. Furthermore, the defense remains effective across both hybrid and fully quantum models, indicating that the robustness improvement is not restricted to a specific architecture or encoding strategy. 

Specifically, this paper makes the following contributions:
\begin{itemize}
    \item We propose controlled passive steering-based state preparation as a practical adversarial defense mechanism for QML models by embedding robustness directly into the quantum encoding stage.
    \item We demonstrate that the proposed method is applicable to both angle encoding and amplitude encoding, making it suitable for both hybrid and fully quantum classifiers.
    \item We evaluate applicability across three datasets (MNIST~\cite{lecun1998mnist}, FashionMNIST~\cite{xiao2017fashionmnist}, and KMNIST~\cite{clanuwat2018deep}), two adversarial attack settings (FGSM and PGD), and three QML models (QNN, QCNN, and VQC) to demonstrate that controlled passive steering-based state preparation provides an effective solution for improving adversarial robustness.
\end{itemize}

The remainder of this paper is organized as follows. Section~\ref{sec:background} presents the necessary background and surveys related efforts. Section~\ref{sec:method} describes the proposed controlled measurement-induced steering-based state preparation framework for replacing the conventional quantum encoding stage. Section~\ref{sec:results} presents the experimental results. Finally,  Section~\ref{sec:conclusion} concludes the paper.






\section{Background and Related Work} \label{sec:background}
In this section, we first provide the necessary background on quantum machine learning and adversarial attacks. Next, we survey related efforts and their limitations to highlight the novelty of our contributions.

\subsection{Quantum Machine Learning} \label{sec:background_qml}

QML incorporates quantum state preparation, quantum circuit transformations, and measurement into learning pipelines for tasks such as feature extraction and classification. In this paper, we consider three representative QML architectures: the quanvolutional neural network (QNN), the quantum convolutional neural network (QCNN), and the variational quantum classifier (VQC).


\subsubsection{Quanvolution Neural Network (QNN)}

\begin{figure}[h]
    \centering
    \includegraphics[width=\linewidth]{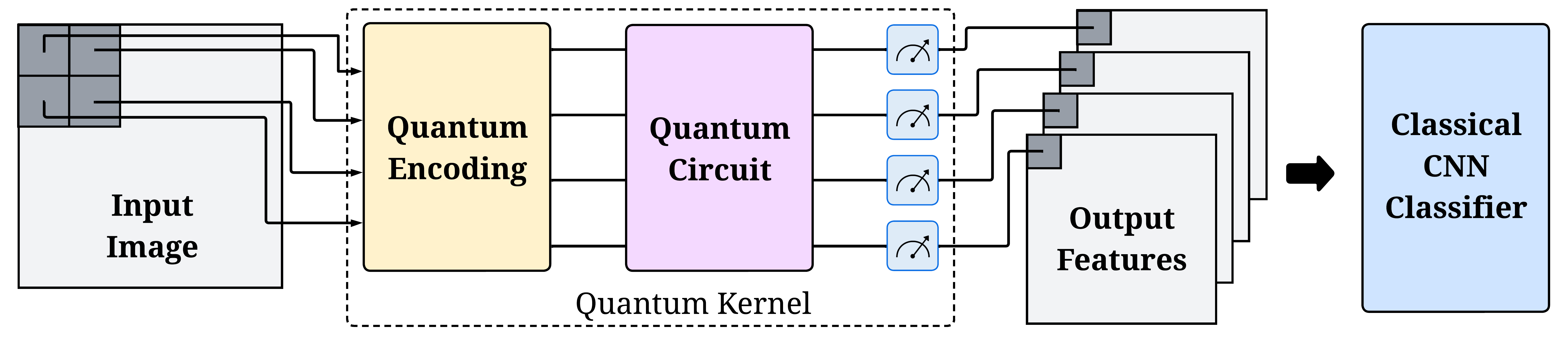}
    \caption{Quanvolution Neural Network (QNN)}
    \label{fig:qnn}
\end{figure}

QNN is a hybrid quantum-classical model in which local patches of the input are processed by quantum circuits acting as quantum filters~\cite{henderson2020quanvolutional}. Similar to a classical convolutional layer, this quanvolutional operation produces feature maps that are subsequently passed to classical layers for downstream learning and classification, as shown in Figure~\ref{fig:qnn}. 

\subsubsection{Quantum Convolutional Neural Networks}

\begin{figure}[h]
    \centering
    \includegraphics[width=1\linewidth]{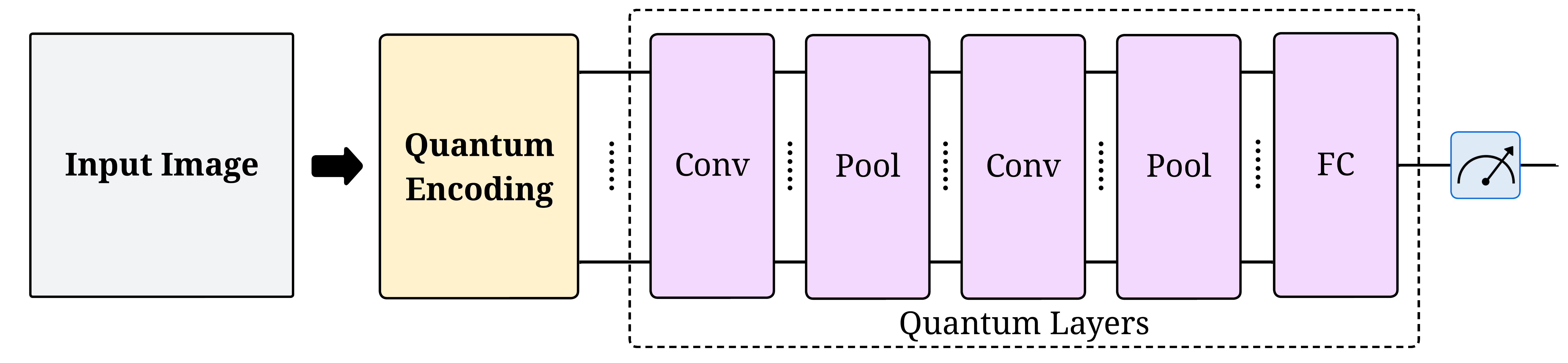}
    \caption{Quantum Convolutional Neural Network (QCNN)}
    \label{fig:qcnn}
\end{figure}

As shown in Figure~\ref{fig:qcnn}, QCNN extends the convolution-and-pooling principle to quantum circuits by applying structured multi-qubit operations in a hierarchical manner~\cite{cong2019quantum}. This architecture enables progressive feature extraction and dimensionality reduction in the quantum domain while using a relatively compact set of trainable parameters.

\subsubsection{Variational Quantum Classifier}

\begin{figure}[h]
    \centering
    \includegraphics[width=1\linewidth]{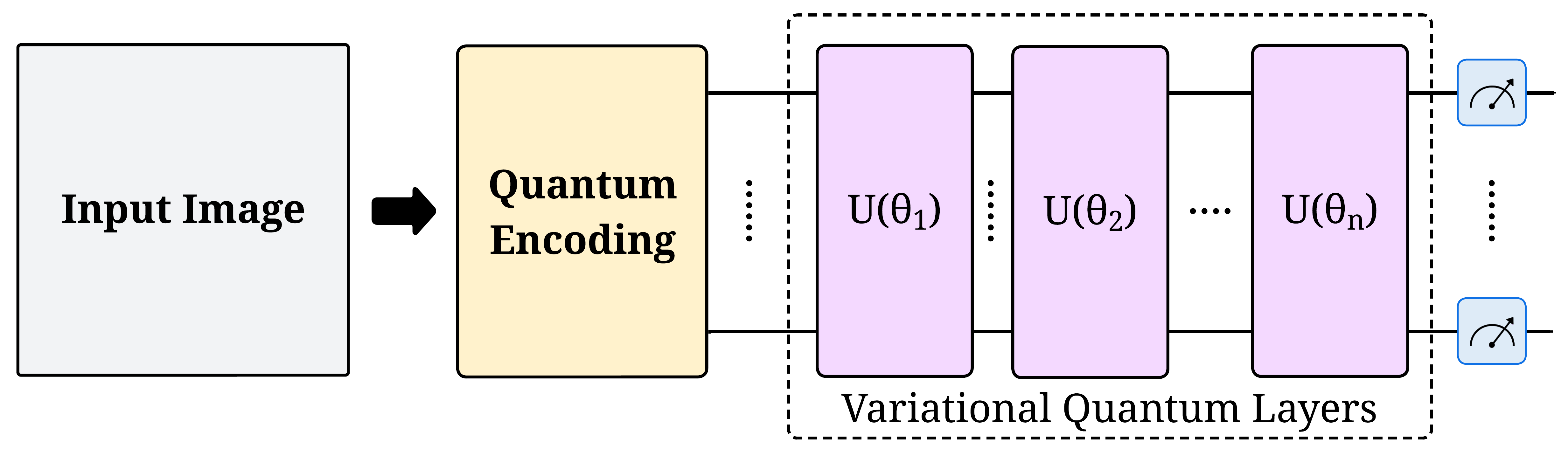}
    \caption{Variational Quantum Classifier (VQC)}
    \label{fig:vqc}
\end{figure}

VQC employs a parameterized quantum circuit as the trainable model for supervised classification~\cite{schuld2020circuit}. Figure~\ref{fig:vqc} depicts the abstract architecture of a VQC. Classical features are first encoded into a quantum state, after which variational single- and two-qubit gates transform the state. The final prediction is then obtained by measuring selected observables.

\subsection{Adversarial Attacks} \label{sec:background_attack}


Adversarial attacks evaluate the robustness of a classifier by adding carefully designed perturbations to the input so that the prediction changes with minimal perturbations in inputs. In classification, given an input $\mathbf{x}$ with label $y$ and a model with loss function $\mathcal{L}(\theta,\mathbf{x},y)$, an adversarial example $\mathbf{x}^{adv}$ is generated by perturbing the original input within a bounded norm constraint. Among the most widely used attacks are the fast gradient sign method (FGSM) \cite{goodfellow2015explaining} and projected gradient descent (PGD) \cite{madry2019deep}.

FGSM is a single-step gradient-based attack that perturbs the input in the direction of the sign of the loss gradient:
\begin{equation}
\mathbf{x}^{adv} = \mathbf{x} + \epsilon \cdot \mathrm{sign}\!\left(\nabla_{\mathbf{x}} \mathcal{L}(\theta,\mathbf{x},y)\right),
\end{equation}
where $\epsilon$ controls the perturbation magnitude. Although simple, FGSM is effective at exposing model sensitivity to small input changes. In contrast, PGD is a stronger iterative attack that repeatedly applies small gradient-based updates and projects the perturbed sample back into the allowed $\epsilon$-ball around the original input:
\begin{equation}
\mathbf{x}^{t+1} = \Pi_{\mathcal{B}_{\epsilon}(\mathbf{x})}
\left(
\mathbf{x}^{t} + \alpha \cdot \mathrm{sign}\!\left(\nabla_{\mathbf{x}} \mathcal{L}(\theta,\mathbf{x}^{t},y)\right)
\right),
\end{equation}
where $\alpha$ is the step size and $\Pi_{\mathcal{B}_{\epsilon}(\mathbf{x})}$ denotes projection onto the valid perturbation set.

\begin{figure*}[t]
    \centering
    \includegraphics[width=\linewidth]{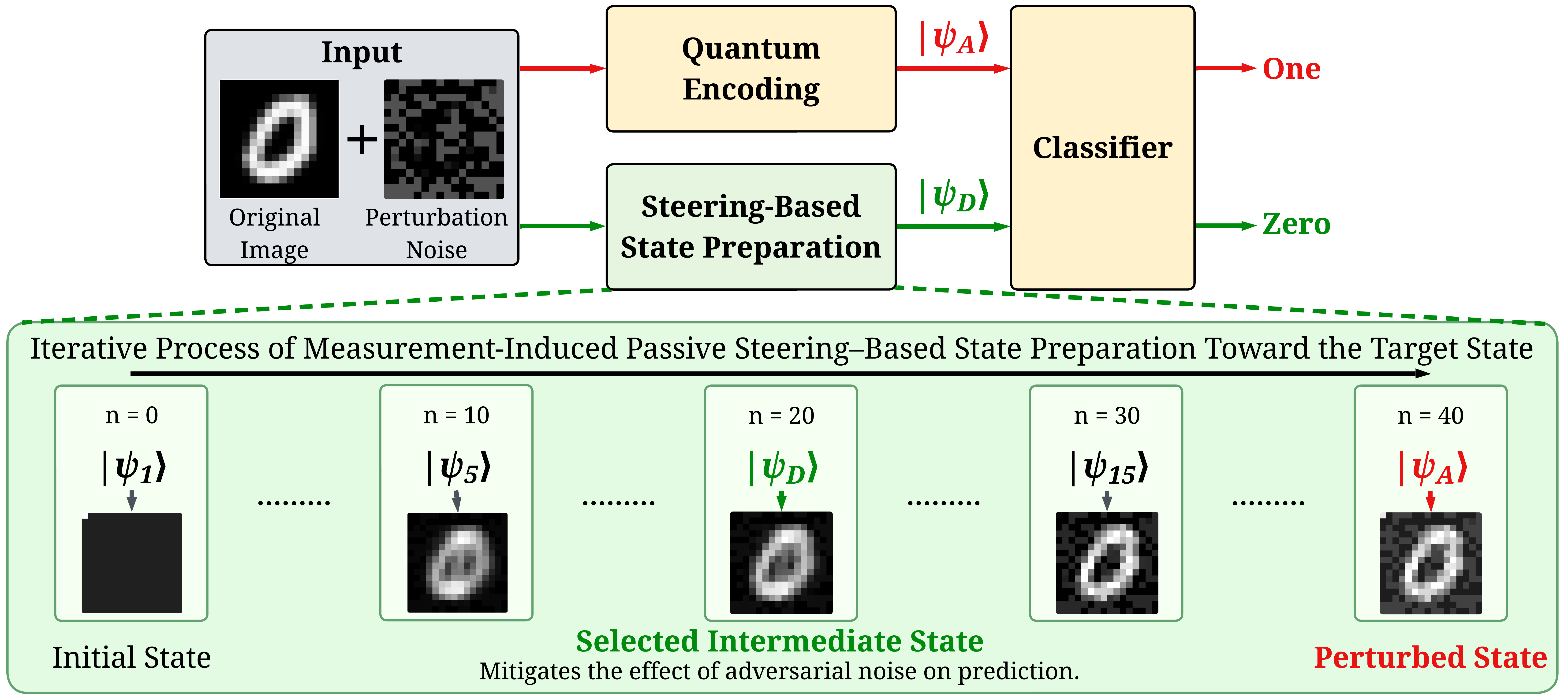}
    \caption{Controlled passive steering–based state preparation at encoding stage as a defense mechanism against adversarial attacks. This illustrates how the conventional quantum encoding stage (red path) is replaced by measurement-induced passive steering (green path). The iterative steering process guides the system toward a controlled intermediate state ($|\psi_D\rangle$), acting as a filter to suppress adversarial perturbation noise before the state is processed by the classifier.}
    \label{fig:overview}
\end{figure*}

\subsection{Related Work} \label{sec:related}

Recent studies have proposed several defense strategies to improve the adversarial robustness of QML models. Adversarial training remains one of the most actively explored approaches. An adversarial training-based defense for quantum classifiers was introduced in~\cite{lu2020quantum}, while~\cite{ren2022experimental} experimentally demonstrated this strategy on a VQC implemented with programmable superconducting qubits. Beyond adversarial training, regularization-based methods have also been investigated to constrain model behavior during training. In particular,~\cite{wendlinger2024comparative} and~\cite{berberich2024training} studied loss-function modifications for robustness enhancement, while~\cite{li2026dual} combined nonlinear quantum encoding with dual regularization to further improve the robustness of QML models.

Other studies have explored robustness enhancement through noise injection and architectural modifications. For example,~\cite{huang2023enhancing} and~\cite{du2021quantum} showed that introducing controlled noise can help suppress adversarial perturbations. Similarly, ~\cite{gong2024enhancing} proposed a random encoding scheme based on unitary transformations or quantum error-correction encoders, that are unknown to the attacker, to defend against man-in-the-middle perturbation injection attacks in the classical communication channel between the sender and the quantum-classifier receiver. Robustness evaluations have also shown that circuit design itself strongly affects vulnerability. In particular, partitioned quantum classifiers~\cite{kananian2025partitioned} and distributed quantum classifiers~\cite{kananian2025adversarial} can become more susceptible to adversarial attacks, revealing a trade-off between scalability and security. Similarly,~\cite{el2024robqunns} demonstrated that the robustness of quantum neural networks depends strongly on the choice of quantum kernel architecture. In a different direction, an input-preprocessing-based defense was demonstrated in~\cite{khatun2025classical}, where a classical autoencoder was employed to preprocess inputs before variational quantum classification.

Although these studies have made important progress, most existing defenses rely on training modifications, such as incorporating adversarial examples into the training set, redesigning the loss function, or adopting specific circuit architectures. Input-preprocessing approaches based on classical autoencoders also require separate training of the preprocessing model, which increases the overall computational overhead. \textit{To the best of our knowledge, no prior work has explored controlled state preparation, achieved through passive steering, as a defense against adversarial attacks in QML systems.} 

Our proposed work investigates replacing the conventional, non-trainable encoding circuit with controlled passive steering as a defense at the quantum state-preparation stage. As a result, with the proposed defense method, the downstream trainable model parameters remain unchanged, while the encoding state preparation stage is modified. Also, our proposed method serves as both the encoding mechanism and an input preprocessing stage. This provides a fundamentally different approach to improving adversarial robustness while preserving the native QML pipeline.

\section{Defending Quantum Classifiers against Adversarial Perturbations using Passive Steering} \label{sec:method}

Figure~\ref{fig:overview} provides an overview of our proposed defense method that replaces the conventional encoding circuit with a measurement-induced passive steering circuit, thereby enabling controllability over the encoded state preparation. By implementing controlled state preparation through passive steering during the encoding stage, the proposed method can suppress adversarial noise while maintaining high model accuracy. We first discuss quantum state preparation using passive steering. Next, we describe quantum encoding using controlled passive steering.

\subsection{Quantum State Preparation using Controlled Passive Steering}
\label{sec:defense}

Measurement-induced passive steering~\cite{volya2024state, volya2023feedback} is an alternative state preparation method in which a target quantum system is driven toward a desired state through repeated interaction with an ancilla, followed by ancilla measurement and reset. Unlike conventional state preparation approaches that assume the availability of a known fiducial state and then apply a calibrated sequence of gates, passive steering uses measurement back-action itself as the mechanism for controlled evolution. In each iteration, the ancilla is prepared in a known state, entangled with the system through a carefully designed unitary operator, measured, and then reinitialized for reuse. Although the measurement is performed only on the ancilla, the resulting back-action on the entangled system gradually steers the system state toward a designated target state. This makes passive steering particularly attractive as a potential alternative state preparation strategy, since the control objective is achieved through a repeated entanglement-measurement-reset cycle rather than through direct initialization of the full system.

Passive steering has been utilized in various contexts, including state preparation for variational quantum eigensolver (VQE) frameworks~\cite{sanjaya2024variational}. In this setting, it can either prepare the reference state before a conventional ansatz or replace the ansatz-based trial-state preparation stage entirely, thereby reducing the dependence on known fiducial initialization and providing a flexible approach to state preparation in noisy variational algorithms.

\begin{figure}[ht]
    \centering
    \includegraphics[width=\linewidth]{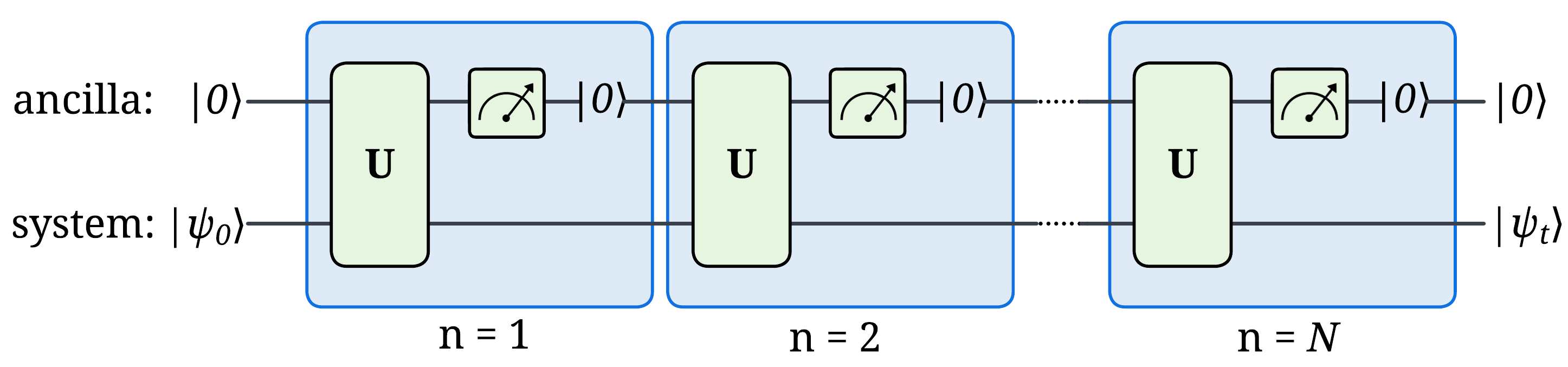}
    \caption{Passive steering circuit}
    \label{fig:steering_circuit}
\end{figure}

As shown in Figure~\ref{fig:steering_circuit}, the passive steering protocol proceeds iteratively through the following steps:
\begin{enumerate}
    \item \textbf{Ancilla-system coupling:} A unitary operator \(U\) is applied to the ancilla and system jointly, creating entanglement between them. If the ancilla is initialized in state \(\rho_A\) and the system is in state \(\rho_S^{(n)}\) after \(n-1\) iterations, the joint state after the \(n^{th}\) iteration becomes
    \(
    \rho_{A\text{-}S}^{(n+1)} = U(\rho_A \otimes \rho_S^{(n)})U^\dagger.
    \)
    This entangling operation is the key step that enables the subsequent measurement on the ancilla to influence the system state.

    \item \textbf{Ancilla measurement:} After the entangling operation, the ancilla is measured. Because the ancilla and system are entangled, this measurement induces a back-action on the system. The updated system state is obtained by tracing out the ancilla,
    \(
    \rho_S^{(n+1)} = \mathrm{Tr}_A[\rho_{A\text{-}S}^{(n+1)}].
    \)
    When the steering operator is properly designed, this measurement-induced evolution increases the overlap between the system state and the target state. 
    
    \item \textbf{Ancilla reset:} After measurement, the ancilla is reinitialized to its original known state so that it can be reused in the next iteration. This reset step makes the protocol iterative and effectively memoryless from the ancilla side, allowing the same control cycle to be applied repeatedly. 

    \item \textbf{Repeated steering toward the target state:} The above cycle is repeated until the system approaches the desired target state \(|\psi_{\oplus}\rangle\). The unitary operator \(U\) is chosen such that the target-state overlap satisfies
    \(
    \langle \psi_{\oplus}|\rho_S^{(n+1)}|\psi_{\oplus}\rangle
    \ge
    \langle \psi_{\oplus}|\rho_S^{(n)}|\psi_{\oplus}\rangle.
    \)
    Therefore, each round of unitary coupling, measurement, and reset incrementally improves the fidelity of the prepared state. 
\end{enumerate}
Two parameters play a central role in this process: the steering strength ($J$) and the number of iterations ($N$). In the state-preparation formulation, the steering unitary is commonly expressed as \(U=\exp(-iJH\delta t)\), where the coupling parameter \(J\) determines the strength of the ancilla-system interaction. Intuitively, \(J\) controls how strongly each application of the protocol pushes the system toward the target state, while the iteration count \(N\) determines how many times the steering cycle is repeated. In practice, these two quantities jointly determine the convergence behavior of the protocol. A larger steering strength can accelerate convergence, whereas increasing the number of iterations generally improves the target-state fidelity, subject to hardware noise and implementation overhead. 

\begin{figure}[htp]
    \centering
    \includegraphics[width=0.8\linewidth]{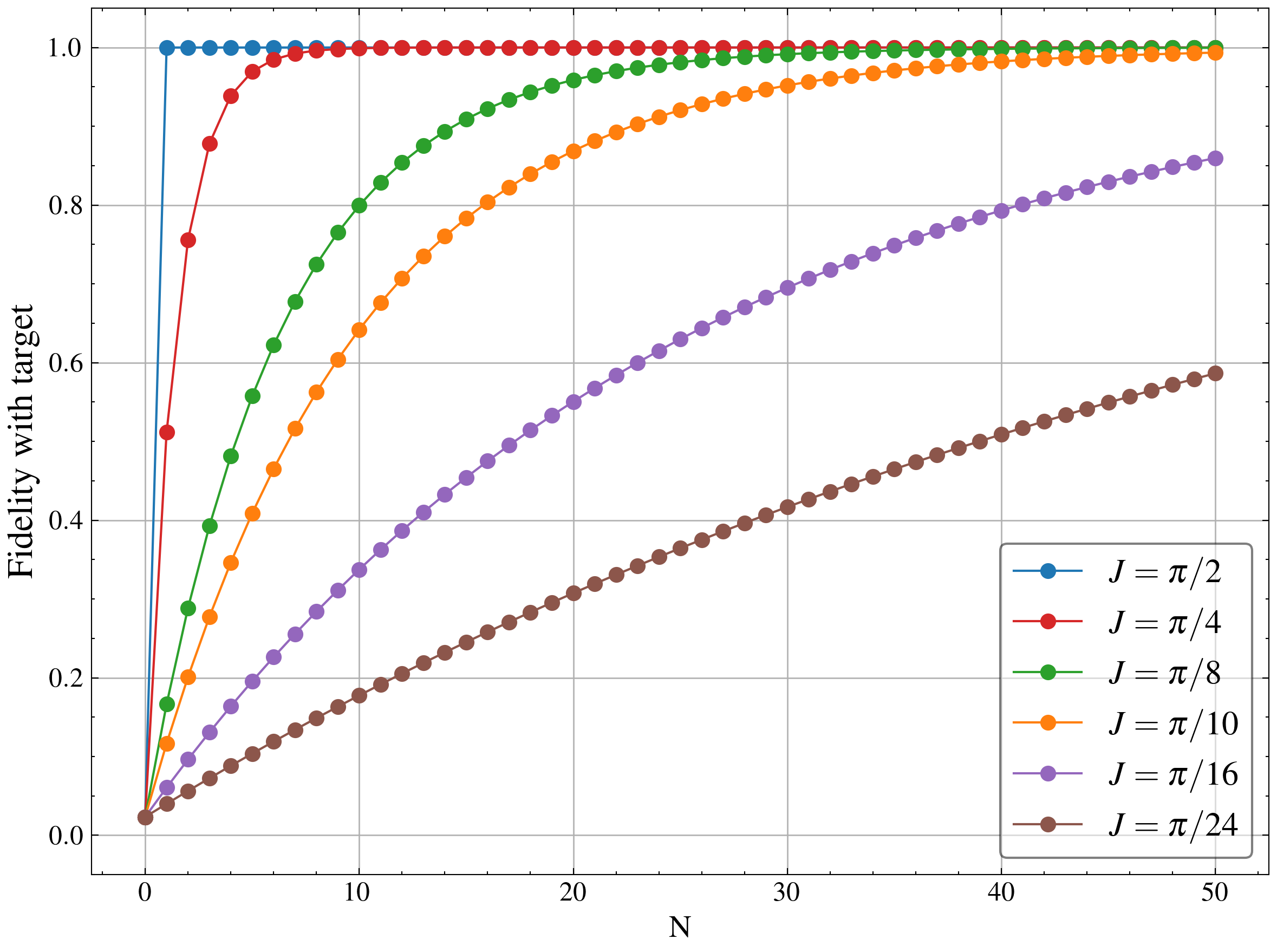}
    \caption{State-preparation fidelity using passive steering for different steering strengths (\(J\)) and numbers of iterations (\(N\)).}
    \label{fig:steering_fidelity}
\end{figure}

Since passive steering-based state preparation can be tuned through the steering strength (\(J\)) and the number of steering iterations (\(N\)), Figure~\ref{fig:steering_fidelity} illustrates the effect of these parameters on the prepared state. In particular, ($J=\pi/2$) provides the maximum steering strength, allowing the initial state to reach the exact target state in fewer iterations, whereas lower steering strengths require more iterations to achieve the same target state. Therefore, by selecting an intermediate steering strength, such as ($\pi/10$) or ($\pi/16$), and varying the number of iterations, the fidelity of the prepared state can be controlled.

The main objective of input preprocessing as a defense against adversarial attacks is to suppress the injected adversarial perturbation while maintaining accuracy on clean data and improving accuracy on adversarial examples. To achieve this, we exploit the configurability offered by passive steering. Instead of preparing the encoded quantum state to exactly match the target state, we intentionally prepare a state that is close to the target state. This affects the input in two different ways, depending on whether the input is clean or adversarial.

\subsubsection{Passive steering in the presence of clean inputs}
For a clean input without perturbation, the controlled passive steering-based encoding introduces a slight deviation because it does not produce the exact encoding state. As a result, the model accuracy for clean inputs may decrease slightly. To limit this effect, we select \(J\) and \(N\) such that the clean accuracy reduction remains below 10\%, ensuring that the loss in clean-data accuracy is minimal while the perturbation is sufficiently suppressed to improve adversarial accuracy, as demonstrated in Section~\ref{sec:results}. Since the attack type and its strength are assumed to be unknown, our empirical results indicate that tuning passive steering to achieve a clean accuracy reduction below 10\% provides a good trade-off between clean accuracy loss and adversarial accuracy improvement.

\begin{figure*}[t]
    \centering
    \begin{subfigure}[b]{0.32\textwidth}
        \centering
        \includegraphics[width=\textwidth]{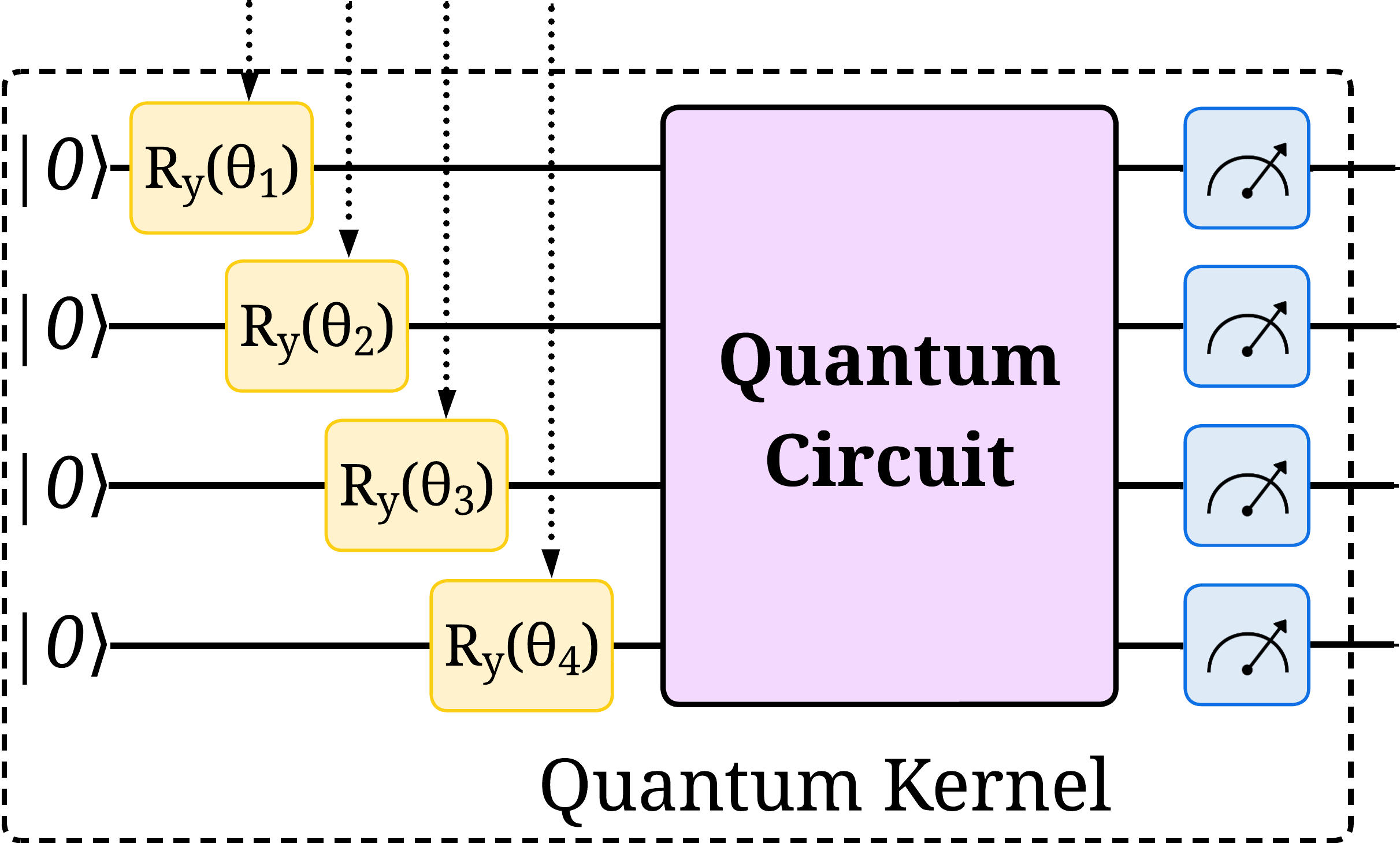}
        \caption{Angle encoding}
        \label{fig:org_angle_enc}
    \end{subfigure}
    \hfill
    \begin{subfigure}[b]{0.32\textwidth}
        \centering
        \includegraphics[width=\textwidth]{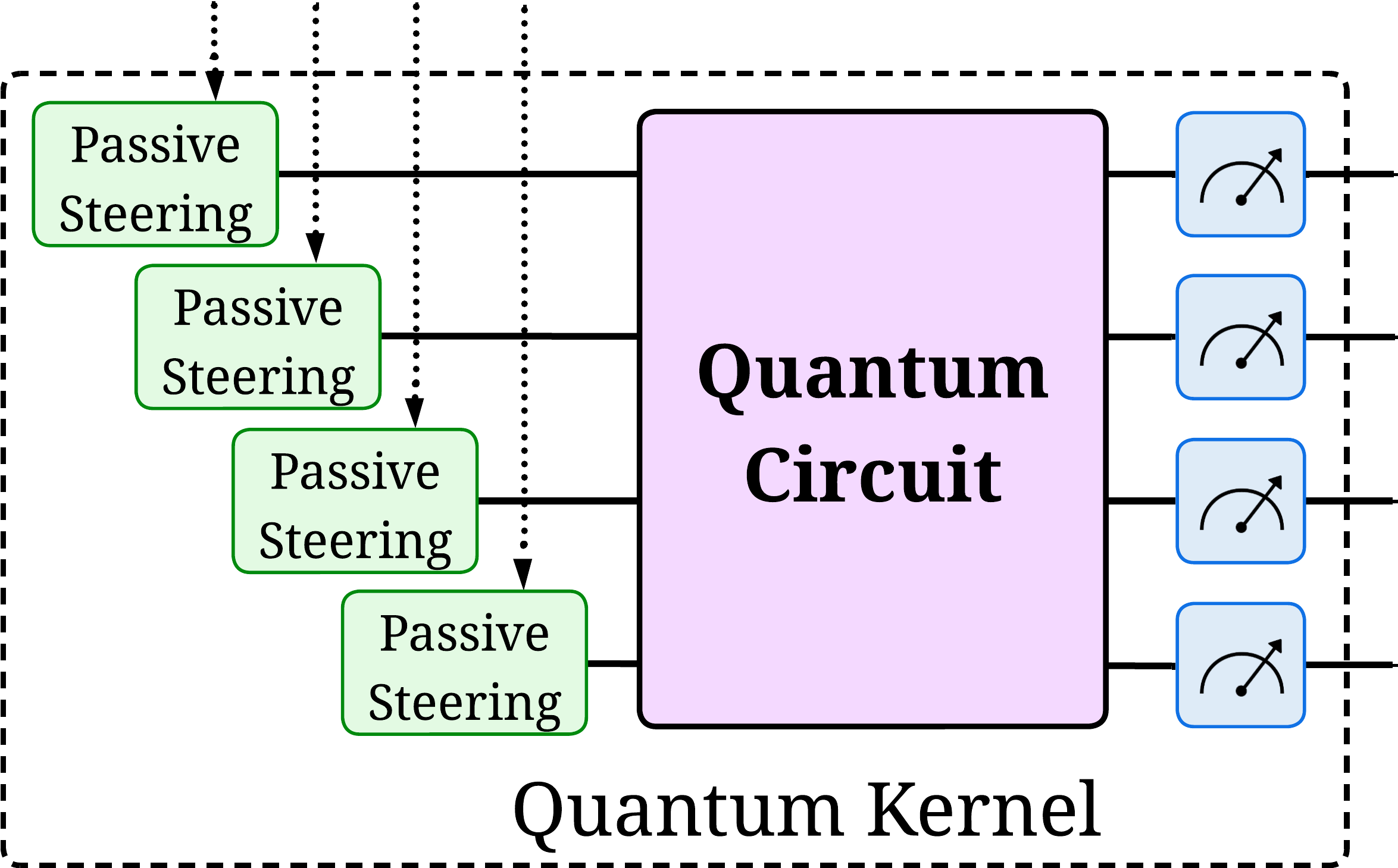}
        \caption{Single-qubit passive steering}
        \label{fig:single_qubit_enc}
    \end{subfigure}
    \hfill
    \begin{subfigure}[b]{0.32\textwidth}
        \centering
        \includegraphics[width=\textwidth]{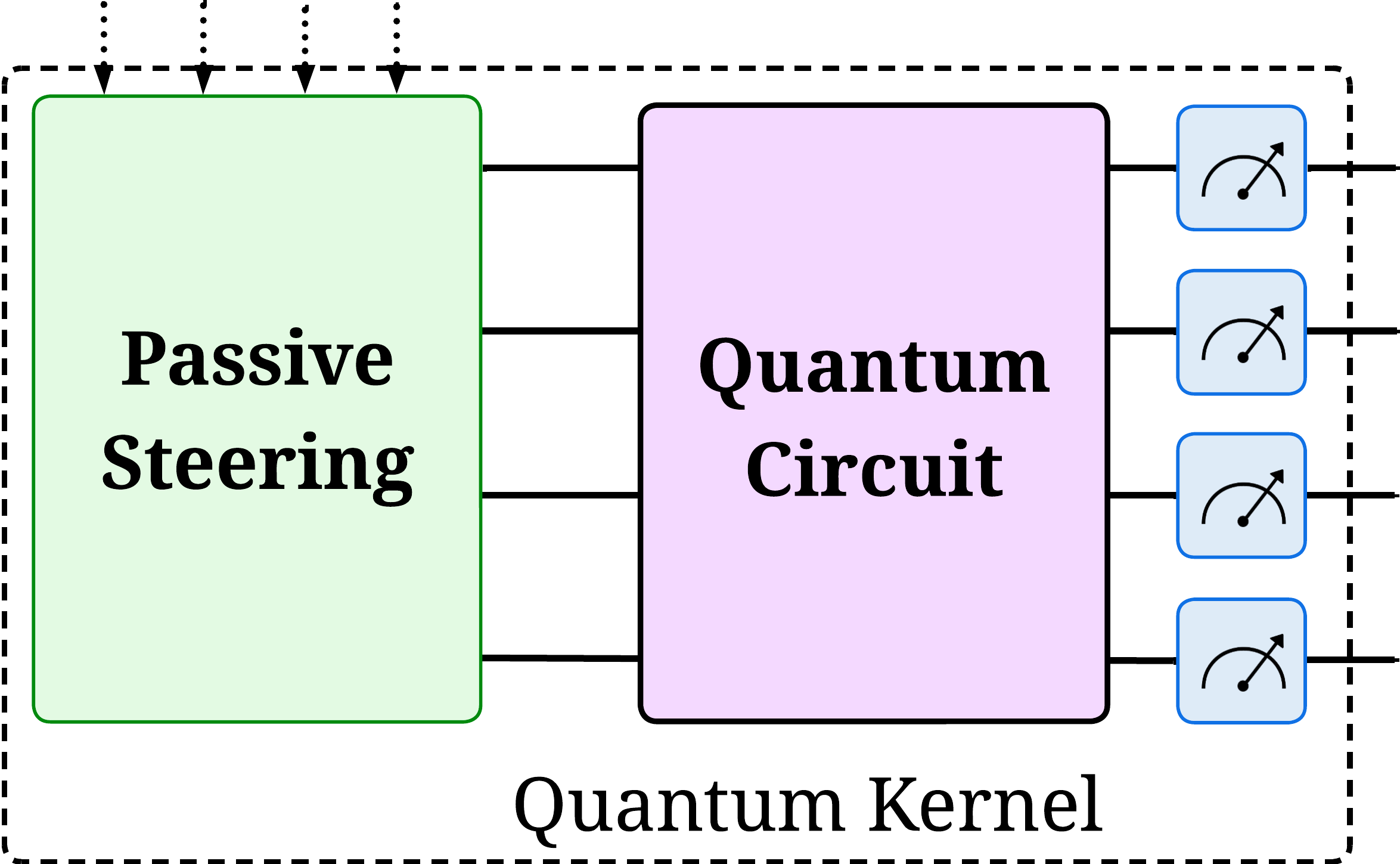}
        \caption{Multi-qubit passive steering}
        \label{fig:multi_qubits_enc}
    \end{subfigure}
    \caption{Quantum kernel circuits used in the QNN. (a) Conventional angle-encoding circuit with single-qubit rotation gates. (b) Proposed defensive angle-encoding circuit with single-qubit controlled passive steering circuits. (c) Proposed defensive angle-encoding circuit with multi-qubit controlled passive steering circuits.}
    \label{fig:}
\end{figure*}


\subsubsection{Passive steering in the presence of adversarial inputs} 
When the input is an adversarial example, the proposed controlled passive steering-based encoding prepares a state with fidelity in the same range relative to the perturbed target state. Due to the fidelity reduction introduced during state preparation, both the original input information and the added perturbation are affected. However, in adversarial attacks, the magnitude of the added perturbation is much smaller than that of the original input. Therefore, the degradation caused by the proposed defense is expected to have a stronger effect on the perturbation than on the underlying input information, which can reduce the influence of adversarial noise on the classifier. This hypothesis is empirically validated in Section~\ref{sec:results} using three datasets under two adversarial attack settings on three different QML models, including one hybrid model and two fully quantum models.

\subsection{Angle Encoding using Passive Steering}
\label{sec:defense_angle}

Quantum algorithms operate on quantum states, whereas most real-world input data are classical. Therefore, a QML model typically begins with an encoding step that maps a classical input vector $x$ to a quantum state \(|\phi(x)\rangle\) in Hilbert space. This mapping, often referred to as a quantum feature map, plays a fundamental role in QML because it determines how information is embedded across qubits during the initial state preparation stage. Among the available encoding strategies, angle encoding is one of the widely used approaches~\cite{khan2024beyond}.

In angle encoding, each feature is embedded into the rotation angle of a single-qubit gate as follows.
\[
|\phi(x)\rangle
=
\bigotimes_{i=1}^{n} R_{\alpha}(x_i)|0\rangle,
\]
where \(R_{\alpha}\) is typically chosen as \(R_x\), \(R_y\), or \(R_z\). For instance, with \(R_y\),
\[
R_y(x_i)|0\rangle
=
\cos\!\left(\frac{x_i}{2}\right)|0\rangle
+
\sin\!\left(\frac{x_i}{2}\right)|1\rangle.
\]
This approach is easy to implement and well-suited to variational circuits, although it usually requires one qubit per feature.

This section explains how angle encoding can be replaced with controlled passive steering-based state preparation. Here, we consider the hybrid QNN model described in Section~\ref{sec:background_qml} as an example. Since angle encoding is applied independently to each qubit, as shown in Figure~\ref{fig:org_angle_enc}, passive steering-based state preparation can be incorporated in two different ways.

\subsubsection{Single-Qubit Steering}

As depicted in Figure~\ref{fig:single_qubit_enc}, in this approach, the steering circuit replaces each individual $R_y$ rotation. This method provides controllability over each encoding channel, although it introduces the cost of repeated circuit execution and extra ancilla qubits. The remaining model flow stays unchanged, as illustrated in Figure~\ref{fig:qnn}. Figure~\ref{fig:qnn_steering_comp} presents an example comparing the quantum kernel outputs obtained using standard $R_y$ rotations and those obtained using passive steering. For adversarial examples, the results clearly show that the controlled passive steering-based encoding suppresses the adversarial perturbation, thereby improving the final classification accuracy.

\subsubsection{Multi-Qubit Steering}

\begin{figure}[ht]
    \centering
    \includegraphics[width=\linewidth]{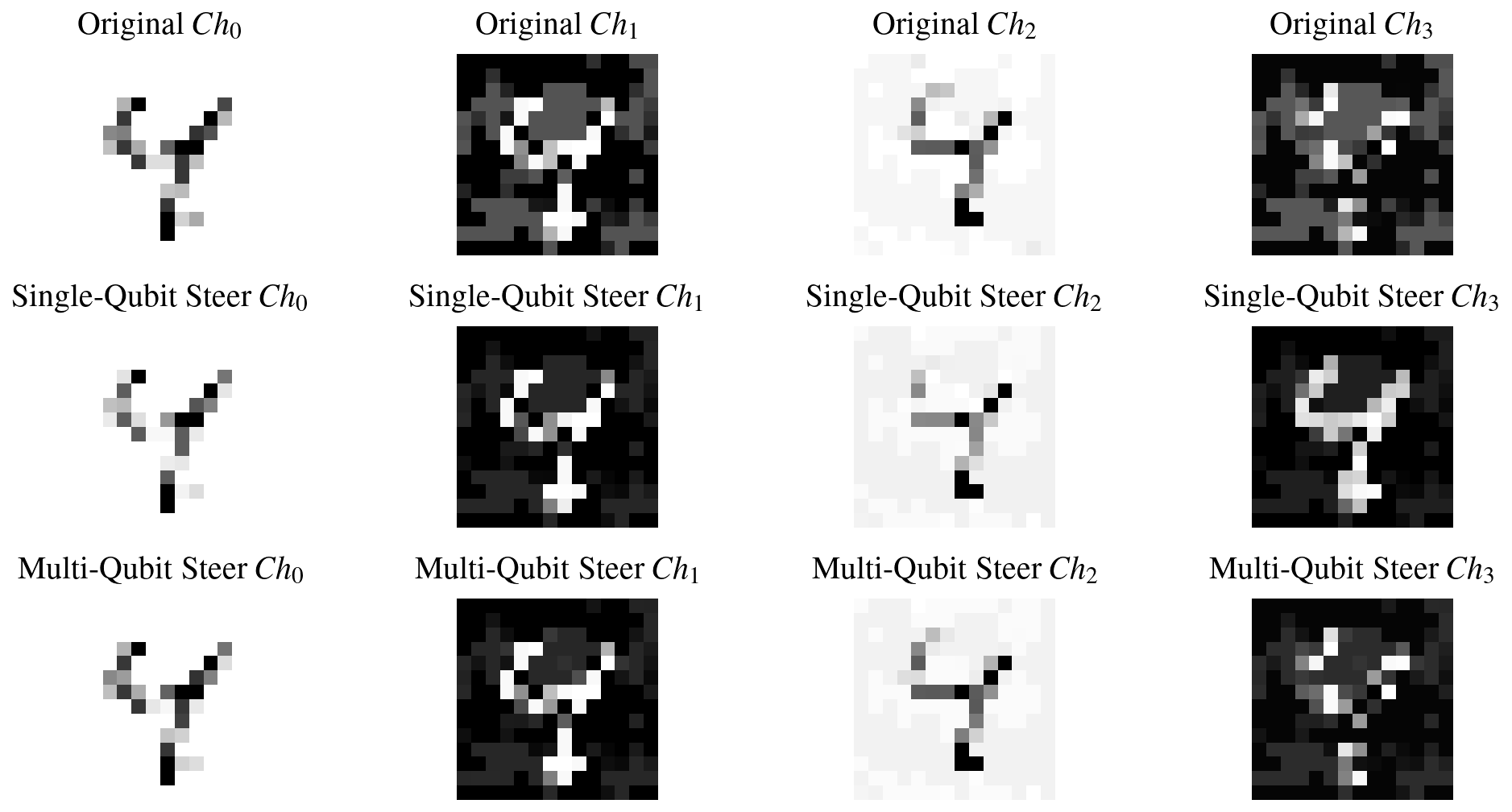}
    \caption{Visual comparison of the outputs produced by standard \(R_y\) rotations, single-qubit controlled passive steering, and multi-qubit controlled passive steering under adversarial input.}
    \label{fig:qnn_steering_comp}
\end{figure}

In this approach, the entire angle-encoding circuit is replaced with a single passive steering circuit, which requires only one ancilla qubit. Figure~\ref{fig:multi_qubits_enc} presents the implementation. Figure~\ref{fig:qnn_steering_comp} illustrates an example comparing the outputs of the conventional angle-encoding stage with those of the multi-qubit controlled passive steering-based encoding under an adversarial attack setting. This figure presents the channel outputs for the same $J$ and $N$ values used in the single-qubit setting. Compared with single-qubit steering, multi-qubit steering suppresses less noise for the same $N$. Therefore, a higher $N$ is required to achieve performance comparable to that of single-qubit steering.

However, single-qubit steering is only applicable when the encoding is performed independently on each qubit, which is generally the case only for simple angle encoding. More complex encoding methods, such as amplitude encoding, involve circuits that entangle multiple qubits. In such cases, replacing the entire encoding circuit with a single steering circuit becomes the more practical approach.

\subsection{Amplitude Encoding using Passive Steering}
\label{sec:defense_amp}

Amplitude encoding is another widely used quantum encoding method~\cite{khan2024beyond} and, compared to angle encoding, it loads data into the amplitudes of a quantum state as follows.
\[
|\phi(x)\rangle
=
\sum_{i=0}^{N-1} x_i |i\rangle,
\qquad
\sum_{i=0}^{N-1}|x_i|^2=1.
\]
Because an \(n\)-qubit system spans \(2^n\) basis states, this method can encode \(N=2^n\) features using only \(n=\log_2 N\) qubits. For non-normalized input, the following normalization needs to be applied.
\[
\tilde{x}=\frac{x}{\|x\|_2}
\]
This makes amplitude encoding compact in terms of qubit resources, although arbitrary state preparation is generally more demanding than the local rotations used in angle encoding, and the input dimension often needs to match a power of two.

To demonstrate the applicability of the proposed defense to QML models with amplitude encoding, we evaluate QCNN and VQC models using the same defense mechanism. Similar to the multi-qubit steering approach discussed in Section~\ref{sec:defense_angle}, the entire amplitude-encoding circuit is replaced with a controlled passive steering circuit. In this setting, the target state of the steering process is the amplitude-embedded image, which would otherwise have been prepared directly and fed into the QML model. Unlike the hybrid approach, where the features correspond to Pauli-Z measurements obtained after angle embedding, in amplitude embedding the image itself, represented in amplitude form, serves as the feature. The intermediate state in the steering approach can therefore be interpreted as a partially steered version of the amplitude-embedded image. As illustrated in Fig. 4, these images are visualized at an intermediate stage of the steering protocol, and the resulting steered quantum states are then fed into the fully quantum model.

Since the steering evolution is halted before reaching the final target state, the input remains farther from the fully adversarial state. As a result, although the clean accuracy is slightly degraded, the adversarial accuracy improves, and our experimental results support this claim, as demonstrated in Section~\ref{sec:results}.

\section{Experiments} \label{sec:results}

In this section, we present experimental results demonstrating the effectiveness of the controlled measurement-induced passive steering-based state preparation as a defense against adversarial attacks. We first outline the experimental setup. Next, we identify suitable parameters for controlled passive steering. Finally, we present the results for both hybrid and fully quantum QML models.

\begin{figure*}[t]
    \centering
    \begin{subfigure}[b]{0.32\textwidth}
        \centering
        \includegraphics[width=\textwidth]{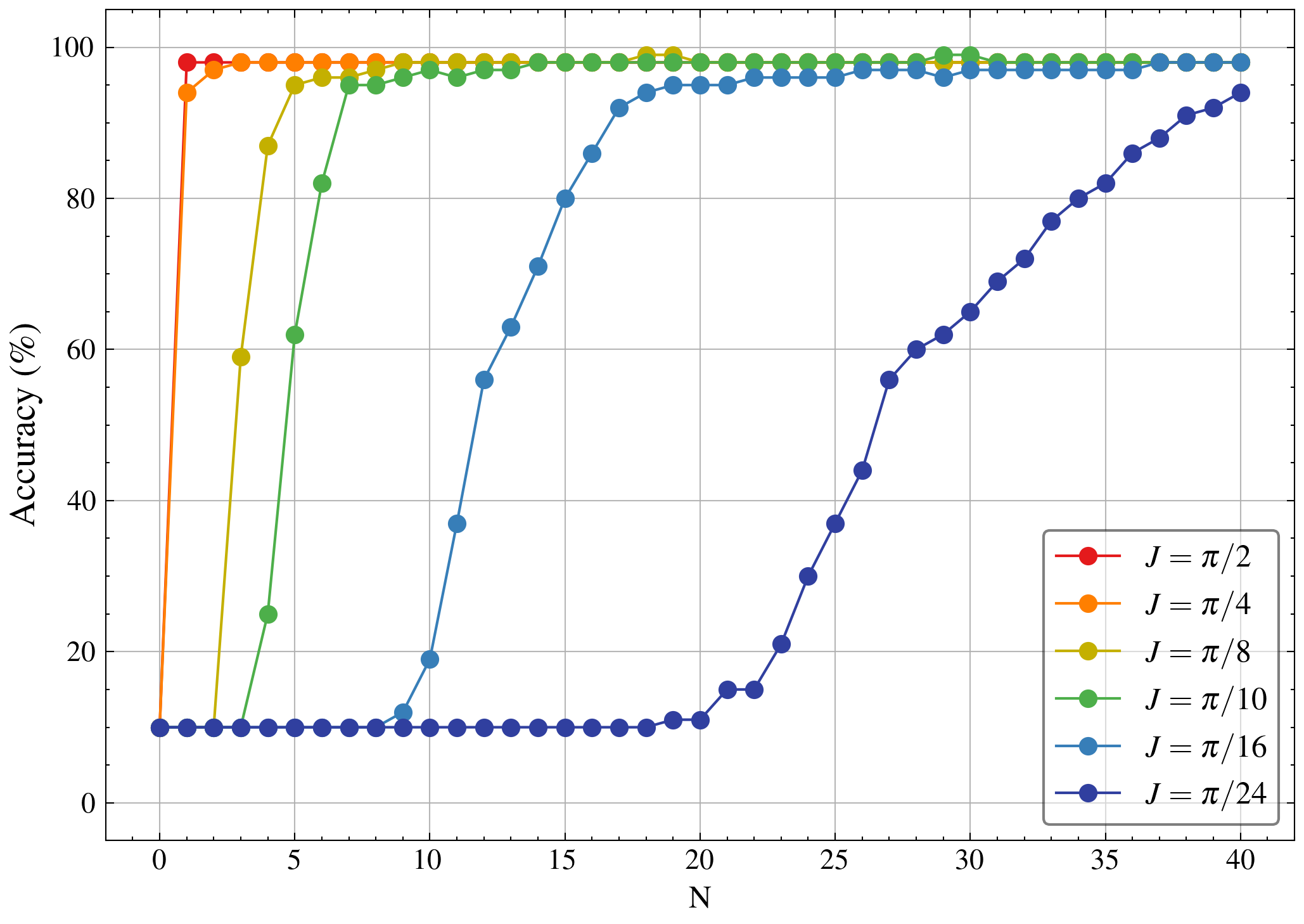}
        \caption{MNIST}
        \label{fig:mnist_clean}
    \end{subfigure}
    \hfill
    \begin{subfigure}[b]{0.32\textwidth}
        \centering
        \includegraphics[width=\textwidth]{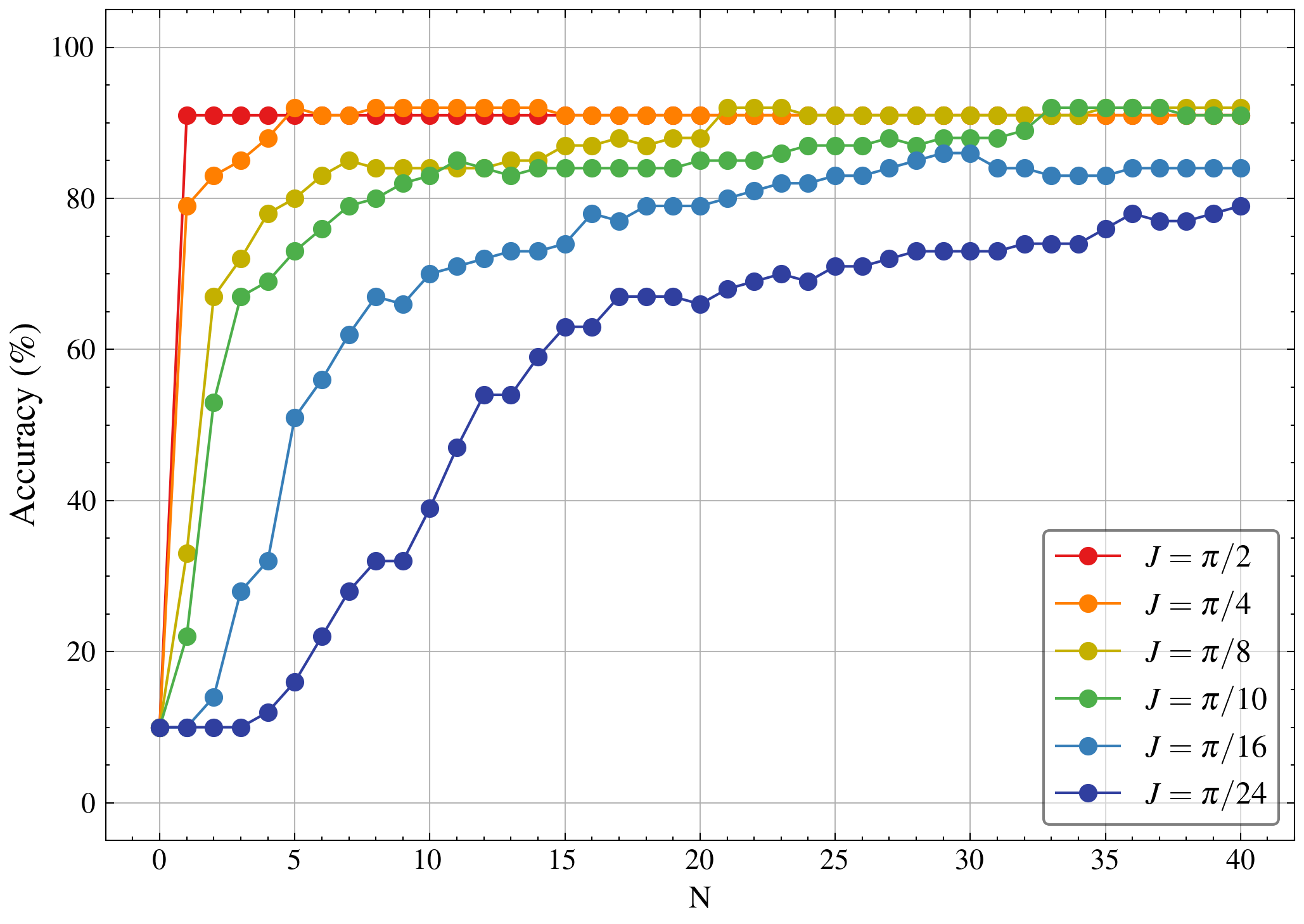}
        \caption{FashionMNIST}
        \label{fig:fmnist_clean}
    \end{subfigure}
    \hfill
    \begin{subfigure}[b]{0.32\textwidth}
        \centering
        \includegraphics[width=\textwidth]{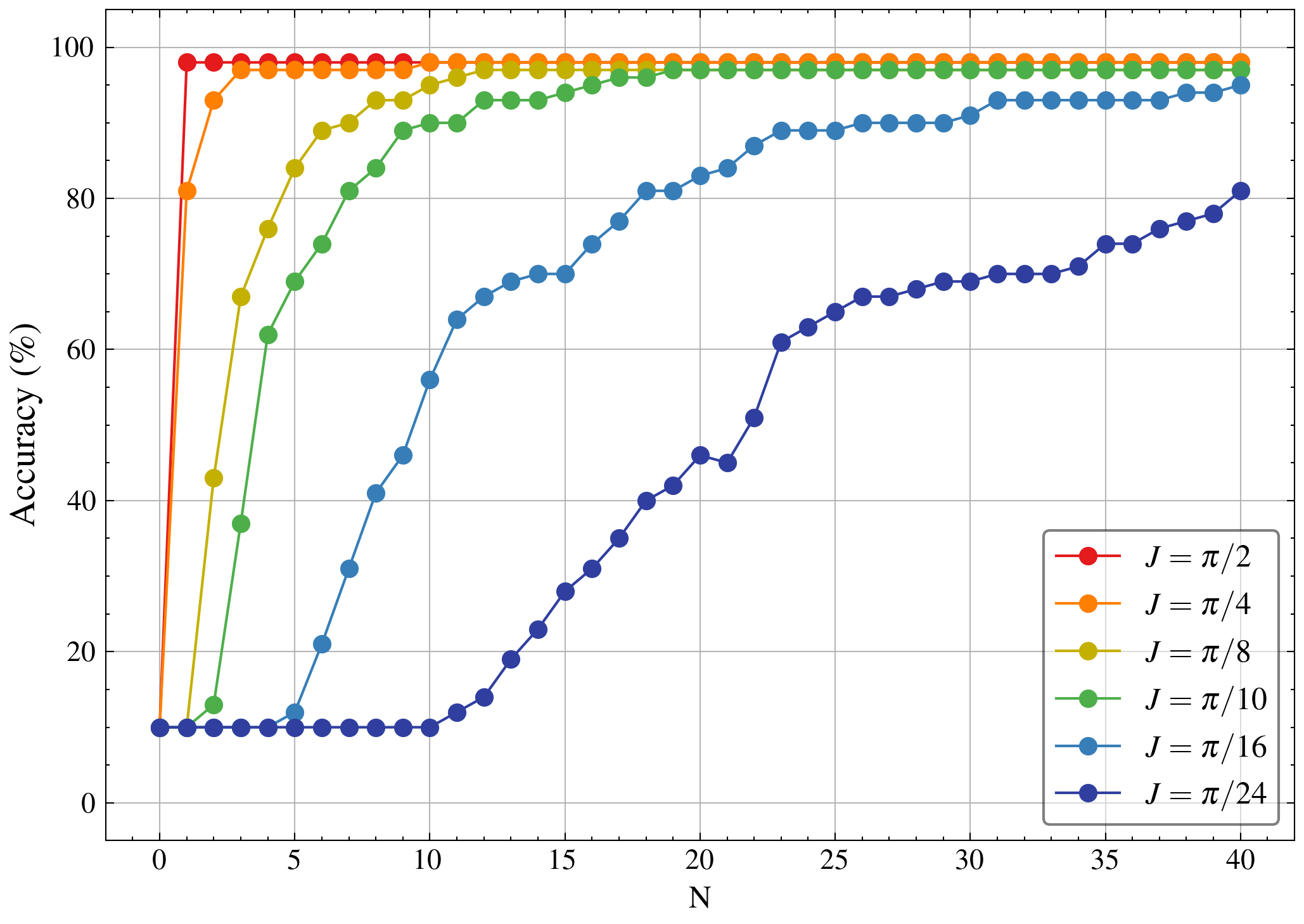}
        \caption{{KMNIST}}
        \label{fig:kmnist_clean}
    \end{subfigure}
    \caption{Clean test accuracy of the QNN with single-qubit controlled passive steering-based encoding for different steering strengths (\(J\)) and numbers of iterations (\(N\)).}
    \label{fig:steering_accuracy}
\end{figure*}

\begin{figure*}[t]
    \centering
    \begin{subfigure}[b]{0.32\textwidth}
        \centering
        \includegraphics[width=\textwidth]{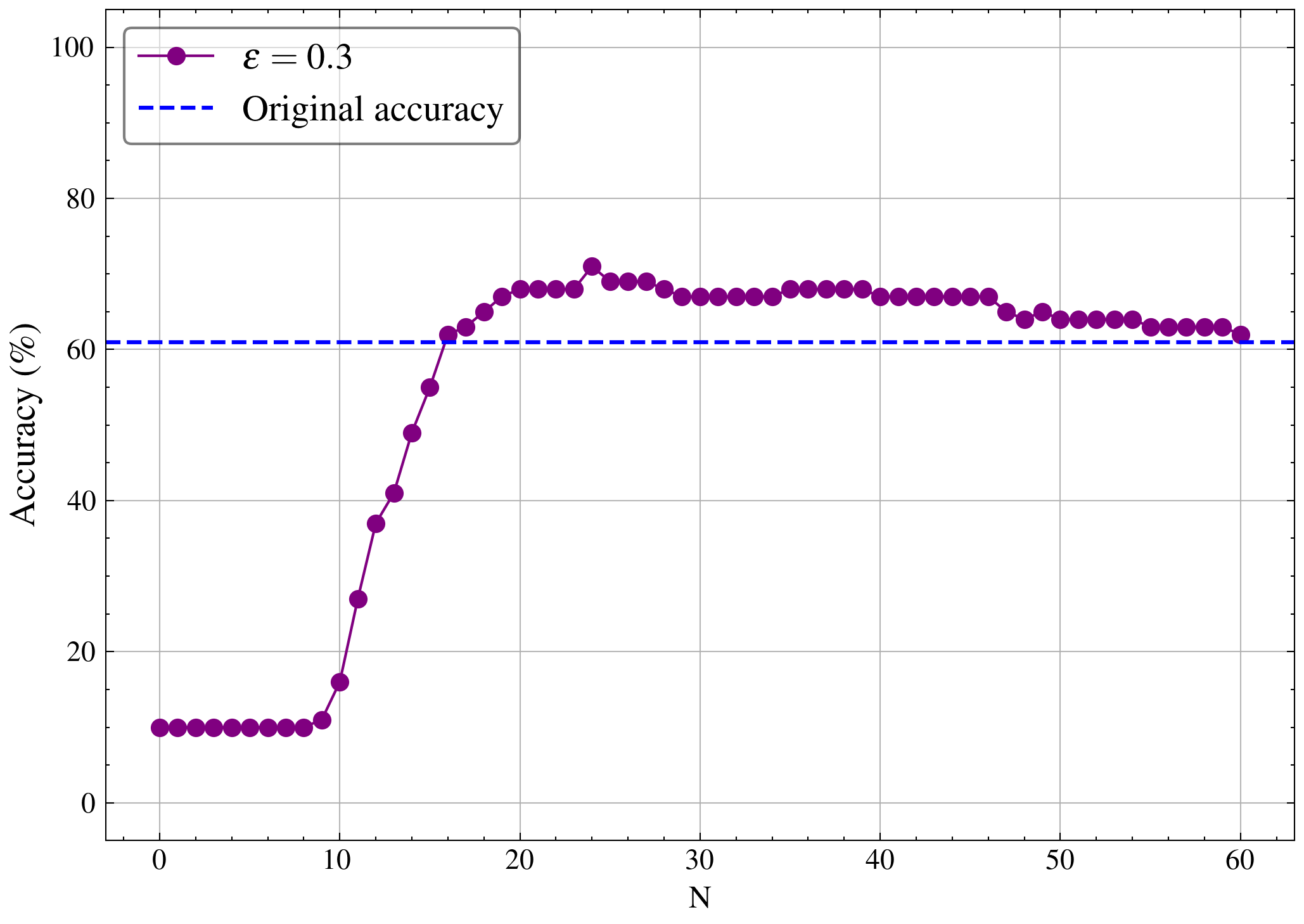}
        \caption{MNIST}
        \label{fig:mnist_attack}
    \end{subfigure}
    \hfill
    \begin{subfigure}[b]{0.32\textwidth}
        \centering
        \includegraphics[width=\textwidth]{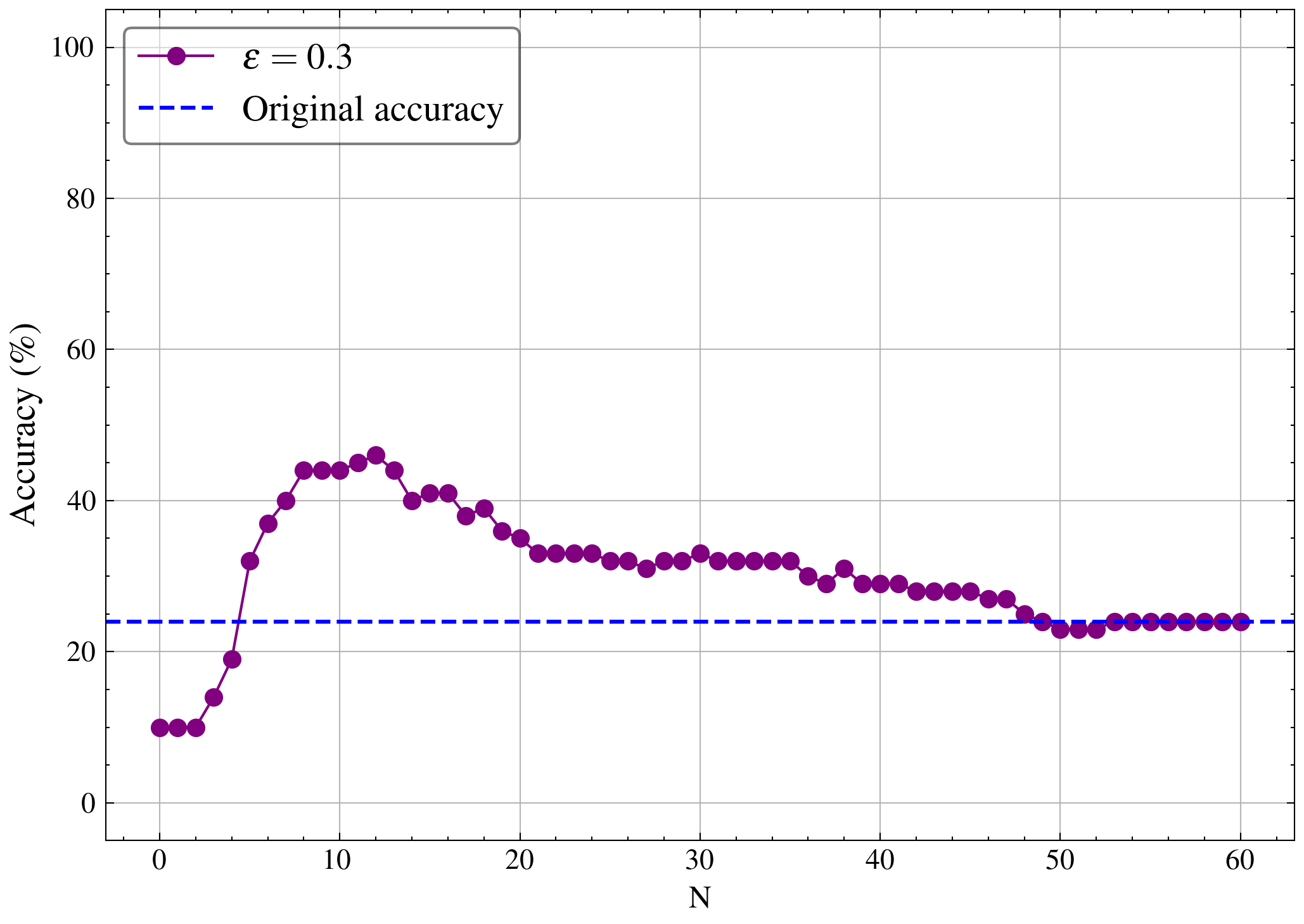}
        \caption{FashionMNIST}
        \label{fig:fmnist_attack}
    \end{subfigure}
    \hfill
    \begin{subfigure}[b]{0.32\textwidth}
        \centering
        \includegraphics[width=\textwidth]{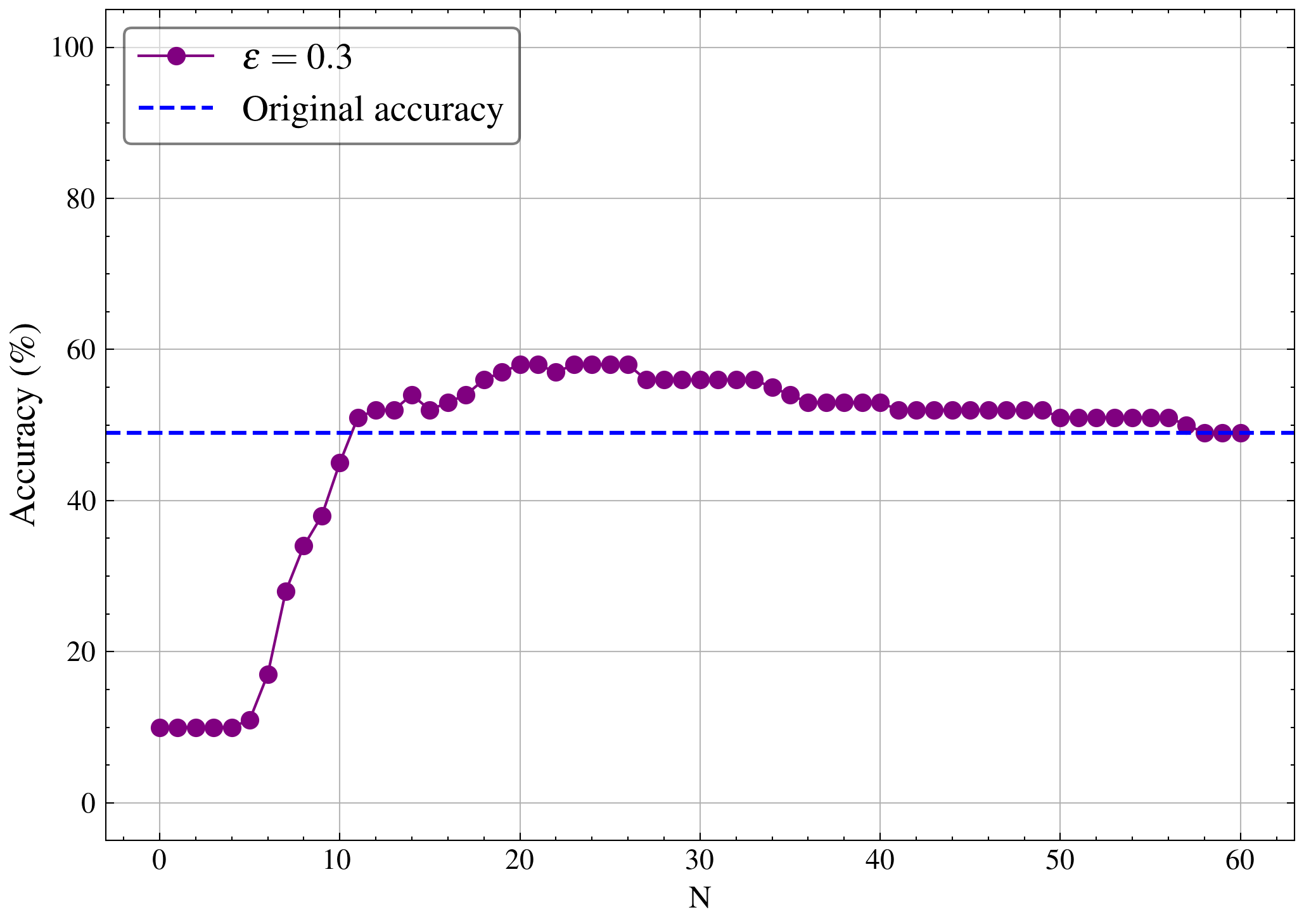}
        \caption{KMNIST}
        \label{fig:kmnist_attack}
    \end{subfigure}
    \caption{Adversarial accuracy under the FGSM attack with perturbation strength \(\epsilon = 0.3\) for the QNN employing single-qubit controlled passive steering-based encoding with steering strength \(J = \pi/16\), shown as a function of the number of iterations \(N\). The blue dashed line represents the adversarial accuracy of the undefended model.}
    \label{fig:steering_adv_accuracy_tuning}
\end{figure*}

\subsection{Experimental Setup}

We conducted our experiments using the PennyLane {v0.44.0}~\cite{bergholm2022pennylane} simulator framework with Python {v3.11.3} and PyTorch {v2.9.1}~\cite{ansel2024pytorch}. The proposed defense mechanism was evaluated using three different QML models, including one hybrid model and two fully quantum models, trained on three datasets and tested under two adversarial attacks. In this section, we first describe the datasets (MNIST, FashionMNIST, and KMNIST). Next we discuss the QML models (QNN, QCNN, and VQC). Finally, we describe the adversarial attacks (FGSM and PGD).

\subsubsection{Datasets}

All three models were trained on the {MNIST}~\cite{lecun1998mnist}, {FashionMNIST}~\cite{xiao2017fashionmnist}, and {KMNIST}~\cite{clanuwat2018deep} datasets. Each dataset contains 70,000 grayscale images of size \(28 \times 28\), with 60,000 training samples and 10,000 testing samples, distributed across 10 classes. MNIST consists of handwritten digits, FashionMNIST contains images of fashion items, and KMNIST includes handwritten Japanese cursive (Kuzushiji) characters.

\subsubsection{Quanvolution Neural Network}

The QNN is implemented as a hybrid quantum-classical model, where a single \(2 \times 2\) quanvolutional layer serves as the initial transformation stage. In this implementation, the \(28 \times 28\) input image is first encoded using angle encoding and then processed by the quanvolutional layer (\texttt{QUANV}), which generates four quantum feature maps of size \(14 \times 14\). The quanvolutional layer is implemented in PennyLane and integrated with PyTorch via the TorchLayer interface. These feature maps are then passed to a classical convolutional backbone consisting of a sequence of convolutional, pooling, and fully connected layers arranged as \texttt{QUANV - CONV1 - POOL1 - CONV2 - POOL2 - FC1 - FC2}. Since this QNN is designed as a multiclass classifier, it is trained and evaluated using the full dataset with all classes. The numbers of training and testing samples used are summarized in Table~\ref{tab:dataset_sizes}. This hybrid architecture enables quantum feature extraction at the input stage while retaining a conventional deep learning pipeline for the final classification task.

\subsubsection{Quantum Convolutional Neural Network}

The QCNN is implemented as a fully quantum model and executed on a PennyLane simulator. Classical inputs are first flattened and encoded into 8 qubits using amplitude encoding. Since amplitude encoding requires an input dimension of \(2^8\), the input images are downscaled to \(16 \times 16\). 

The QCNN consists of three stages of parameterized quantum convolution and pooling layers, which progressively concentrate the encoded information onto the final qubit. Therefore, unlike the QNN, which is implemented as a multiclass classifier, the QCNN is implemented here as a binary classifier. Accordingly, we consider only the first two classes from each dataset. The numbers of training and testing samples used are summarized in Table~\ref{tab:dataset_sizes}. The circuit contains 48 parameters in the convolution layers and 24 parameters in the pooling layers, resulting in a total of 72 trainable parameters, where each two-qubit block is parameterized by three rotation angles. The model outputs the Pauli-\(Z\) expectation value of the final qubit, which is then mapped to a two-dimensional output for binary classification.

\subsubsection{Variational Quantum Classifier}

Since the QCNN is designed to perform binary classification on downscaled \(16 \times 16\) images, the VQC model is configured in a similar manner to enable a fair evaluation of the proposed defense on fully quantum models. Specifically, the VQC encodes the downscaled inputs into 8 qubits using amplitude encoding for binary classification. The encoded state is then processed by three layers of \texttt{StronglyEntanglingLayers}, which provide a general trainable ansatz over the active qubits. The model outputs the Pauli-\(Z\) expectation value of qubit 8, which is subsequently mapped to a two-dimensional output for binary classification.

\begin{table}[htbp]
    \centering
    \caption{Number of training and testing images in MNIST, KMNIST, and FashionMNIST under two dataset configurations: full dataset and subset containing only the first two classes.}
    \label{tab:dataset_sizes}
    \small
    \begin{tabular}{lcccc}
        \toprule
        \multirow{2}{*}{\textbf{Dataset}} 
        & \multicolumn{2}{c}{\textbf{Training Images}} 
        & \multicolumn{2}{c}{\textbf{Testing Images}} \\
        \cmidrule(lr){2-3} \cmidrule(lr){4-5}
        & \textbf{Full} & \textbf{Binary} 
        & \textbf{Full} & \textbf{Binary} \\
        \midrule
        MNIST        & 60{,}000 & 12{,}665 & 10{,}000 & 2{,}115 \\
        KMNIST       & 60{,}000 & 12{,}000 & 10{,}000 & 2{,}000 \\
        FashionMNIST & 60{,}000 & 12{,}000 & 10{,}000 & 2{,}000 \\
        \bottomrule
    \end{tabular}
\end{table}

\subsubsection{Adversarial Attacks}

To evaluate the effectiveness of the proposed defense, we consider two gradient-based adversarial attacks: FGSM and PGD. Adversarial examples are generated in a white-box setting using the original model without defense. In both attacks, the perturbation budget $\epsilon$ is varied to examine the robustness of the defense under different attack strengths. For PGD, we set the step size to $\alpha = 0.02$ and use 20 iterations.

\subsection{Parameter Selection for Controlled Passive Steering}

Before applying the proposed defense by replacing the existing encoding circuit with passive steering, suitable values of \(J\) and \(N\) must be identified, as discussed in Section~\ref{sec:defense}. Because the proposed method prepares a state that is close to, but not identical to, the original encoded quantum state, it inevitably affects both clean and adversarial accuracy. Thus, the goal is to determine a state-preparation configuration that minimizes the loss in clean accuracy while maximizing the model robustness under adversarial attacks.

To limit the impact on clean accuracy, we evaluate the test accuracy of the defended model and select pairs of \(J\) and \(N\) that produce a clean accuracy reduction below 10\%. Our empirical results show that the parameter (\(J\) and \(N\)) selection can be performed using a partial test set, and that the selected parameter pairs remain suitable for the defense. This provides an efficient way to identify appropriate \(J\) and \(N\) values without requiring the full test dataset.
Figure~\ref{fig:steering_accuracy} illustrates the variation in clean accuracy for different \(J\) and \(N\) values on the MNIST, FashionMNIST, and KMNIST datasets using 100 test images, after applying the single-qubit steering as the encoding circuit in the QNN.

Based on Figure~\ref{fig:steering_accuracy}, several observations can be made. First, for a given dataset, the clean accuracy of the defended model generally increases with increasing \(J\) and \(N\). This behavior is expected because, as shown in Figure~\ref{fig:steering_fidelity}, increasing \(J\) and \(N\) increases the fidelity between the target state and the state prepared using passive steering. Second, for a given dataset, there are multiple suitable pairs of \(J\) and \(N\) that satisfy the clean-accuracy constraint. Moreover, even for the same model, the appropriate \(N\) value may vary across datasets for a fixed \(J\). This is because the undefended model exhibits different baseline clean test accuracies on different datasets.

Moreover, Figure~\ref{fig:steering_adv_accuracy_tuning} shows the FGSM attack results at perturbation strength \(\epsilon = 0.3\) for the QNN model with single-qubit passive steering-based encoding using \(J = \pi/16\). As \(N\) increases, the adversarial accuracy first increases up to a certain point and then begins to decrease. This behavior supports the explanation in Section~\ref{sec:defense}, when controlled passive steering is used to prepare an intermediate state along the trajectory toward the exact encoded state, it can suppress adversarial noise and thereby improve adversarial accuracy. However, when \(N\) is increased beyond a specific point, the passive steering circuit produces a state that becomes very close to the exact encoded state, which reduces its ability to suppress adversarial perturbations. This trend can be observed from the drop in adversarial accuracy after a certain range of \(N\) values.

Furthermore, based on Figure~\ref{fig:steering_accuracy}, for example, the range \(N \in [17,30]\) with \(J = \pi/16\) for MNIST yields a clean accuracy reduction below 10\% and Figure~\ref{fig:steering_adv_accuracy_tuning} shows that the same range of \(N\) values provides the highest adversarial accuracy under the FGSM attack. This suggests that selecting a \((J, N)\) pair that yields an acceptable reduction in clean accuracy also leads to a strong improvement in adversarial accuracy. Based on this analysis, we determine the set of \(J\) and \(N\) values using clean accuracy, as summarized in Table~\ref{tab:jn_pairs}. These values are used in the remaining experiments to evaluate the effectiveness of the proposed defense.

\begingroup
\begin{table}[htbp]
\centering
\caption{Selected $(J,N)$ pairs for different models and datasets. QNN (S) is the single-qubit steering setting and QNN (M) is the multi-qubit setting.}
\label{tab:jn_pairs}
\small
\begin{tabular}{lccc}
\toprule
\multirow{2}{*}{\textbf{Model}} & \multicolumn{3}{c}{\textbf{Datasets}} \\
\cmidrule(lr){2-4}
 & \textbf{MNIST} & \textbf{FashionMNIST} & \textbf{KMNIST} \\
\midrule
QNN (S)  & $\rule{0pt}{0ex}(\dfrac{\pi}{16}, 27)$ & $(\dfrac{\pi}{16}, 20)$ & $(\dfrac{\pi}{16}, 24)$ \\
QNN (M)  & $\rule{0pt}{4ex}(\dfrac{\pi}{16}, 40)$ & $(\dfrac{\pi}{16}, 25)$ & $(\dfrac{\pi}{16}, 35)$ \\
QCNN & $\rule{0pt}{4ex}(\dfrac{\pi}{10}, 10)$ & $(\dfrac{\pi}{10}, 12)$ & $(\dfrac{\pi}{10}, 23)$ \\
VQC  & $\rule{0pt}{4ex}(\dfrac{\pi}{10}, 20)$ & $(\dfrac{\pi}{10}, 18)$ & $(\dfrac{\pi}{10}, 15)$ \\
\bottomrule
\end{tabular}
\end{table}
\endgroup

\begin{figure*}[htbp]
    \centering
    \begin{subfigure}{0.32\linewidth}
        \centering
        \includegraphics[width=\linewidth]{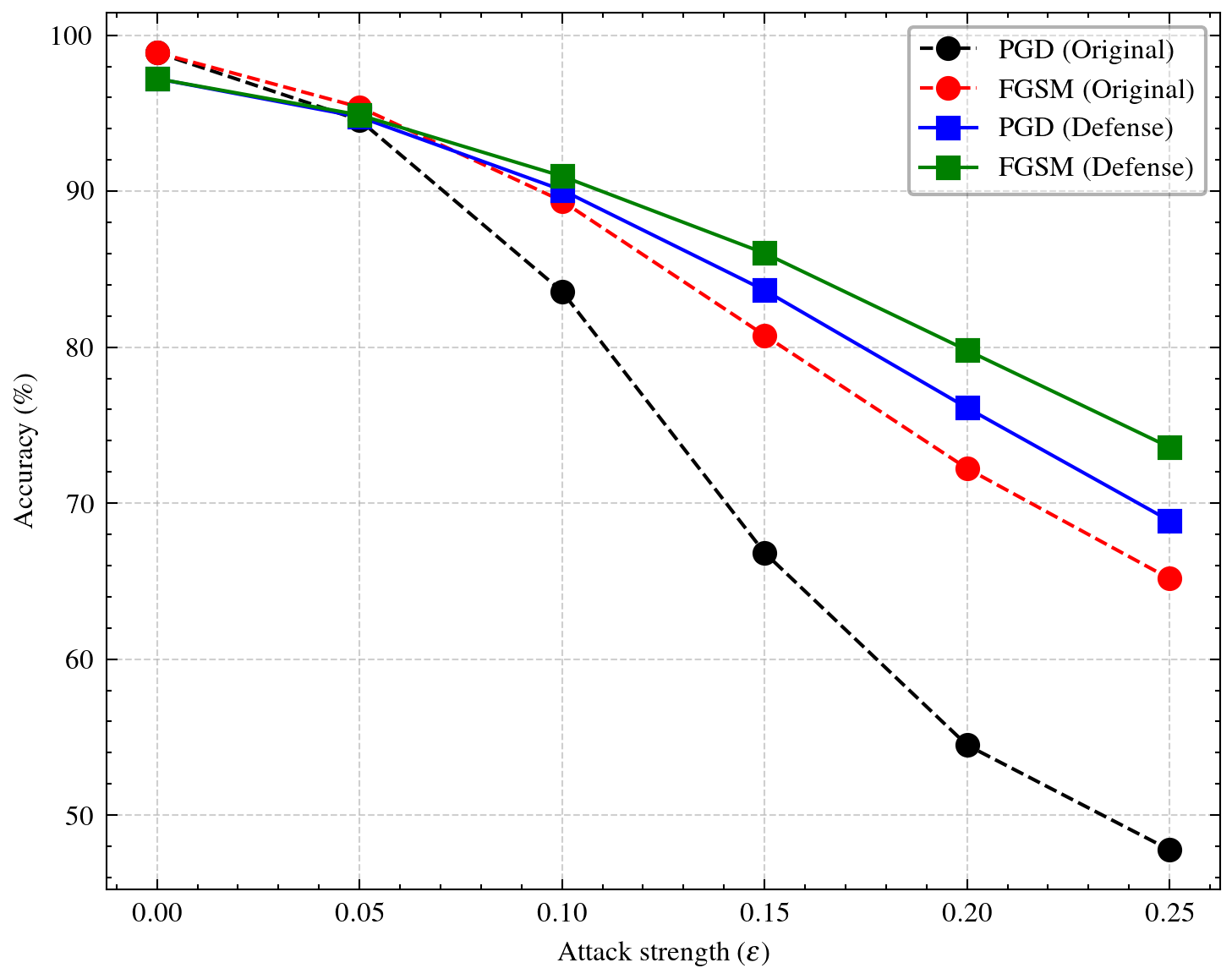}
        \caption{QNN-MNIST (single-qubit)}
        \label{fig:qnn_mnist}
    \end{subfigure}
    \begin{subfigure}{0.32\linewidth}
        \centering
        \includegraphics[width=\linewidth]{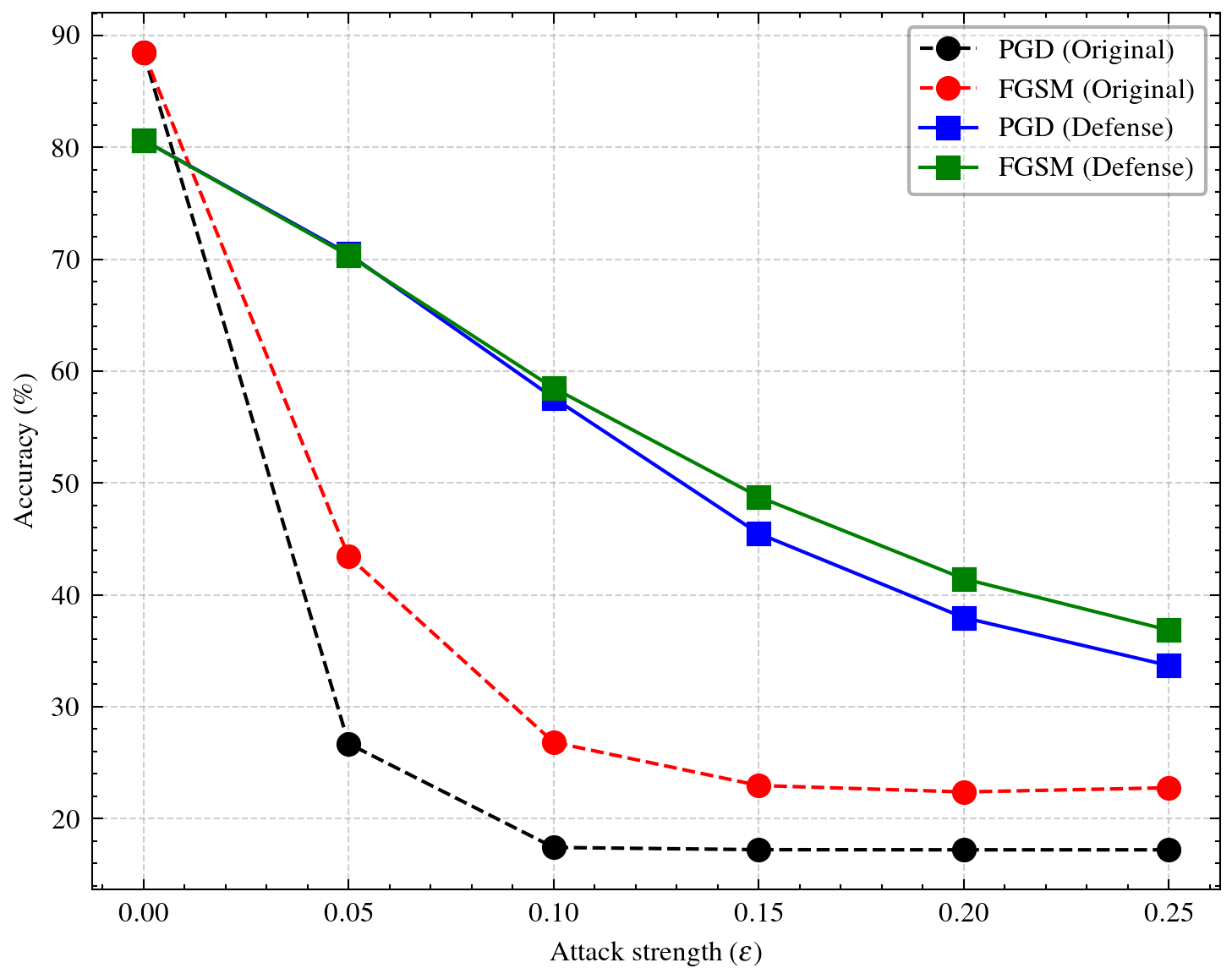}
        \caption{QNN-FashionMNIST (single-qubit)}
        \label{fig:qnn_fashionmnist}
    \end{subfigure}
    \begin{subfigure}{0.32\linewidth}
        \centering
        \includegraphics[width=\linewidth]{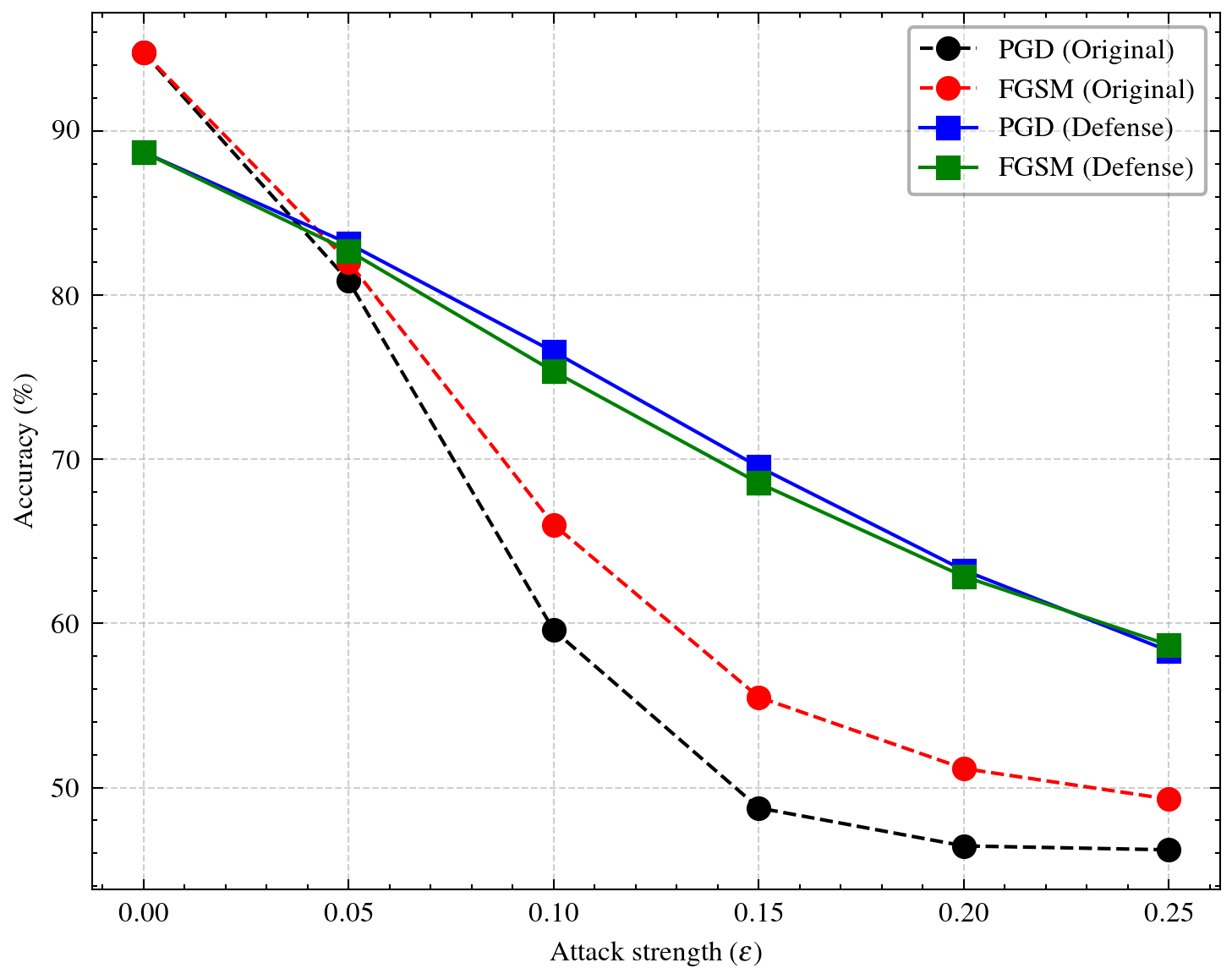}
        \caption{QNN-KMNIST (single-qubit)}
        \label{fig:qnn_kmnist}
    \end{subfigure}

\vspace{0.2in}

    \begin{subfigure}{0.32\linewidth}
        \centering
        \includegraphics[width=\linewidth]{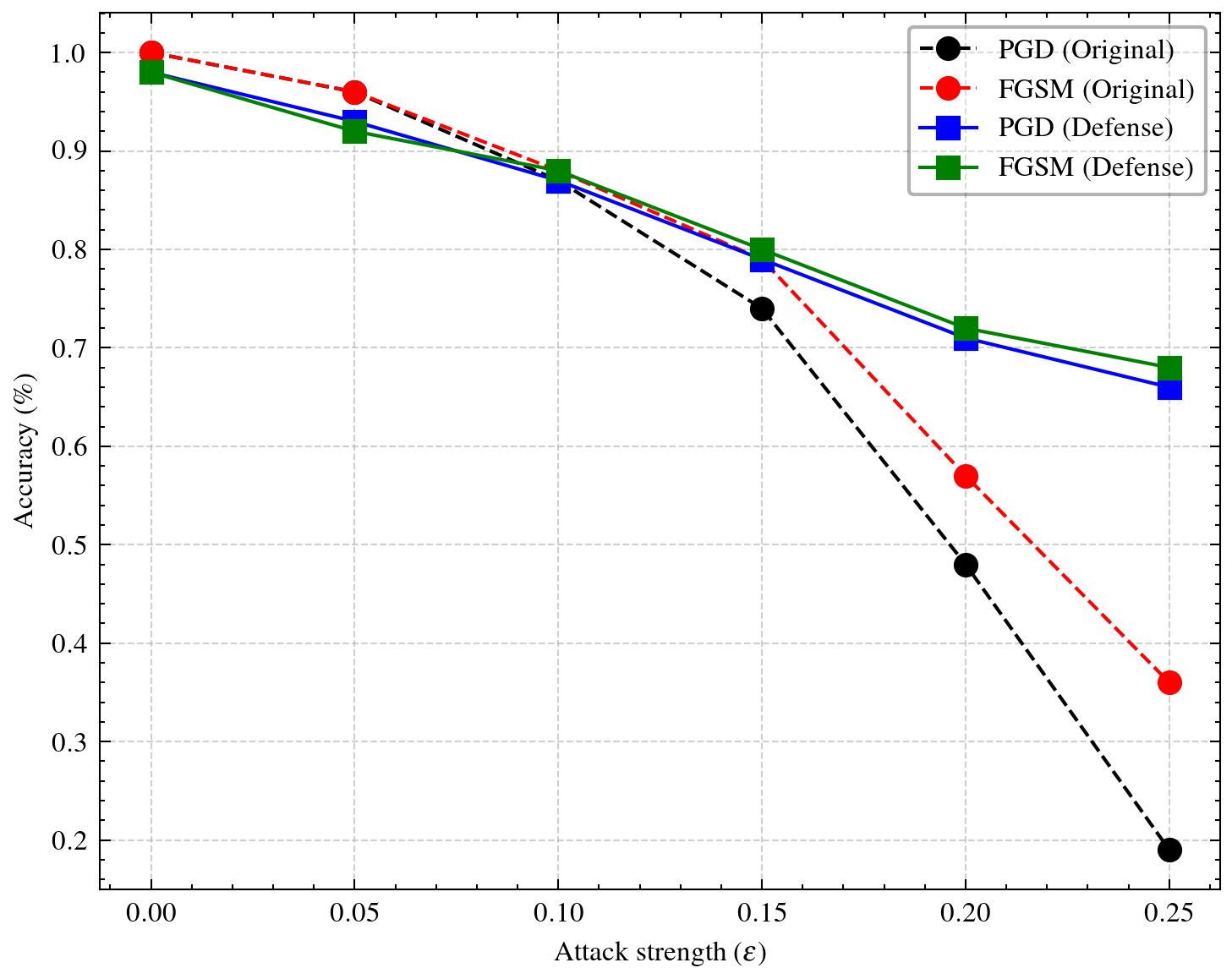}
        \caption{QNN-MNIST (multi-qubit)}
        \label{fig:qnn_mnist)full}
    \end{subfigure}
    \begin{subfigure}{0.32\linewidth}
        \centering
        \includegraphics[width=\linewidth]{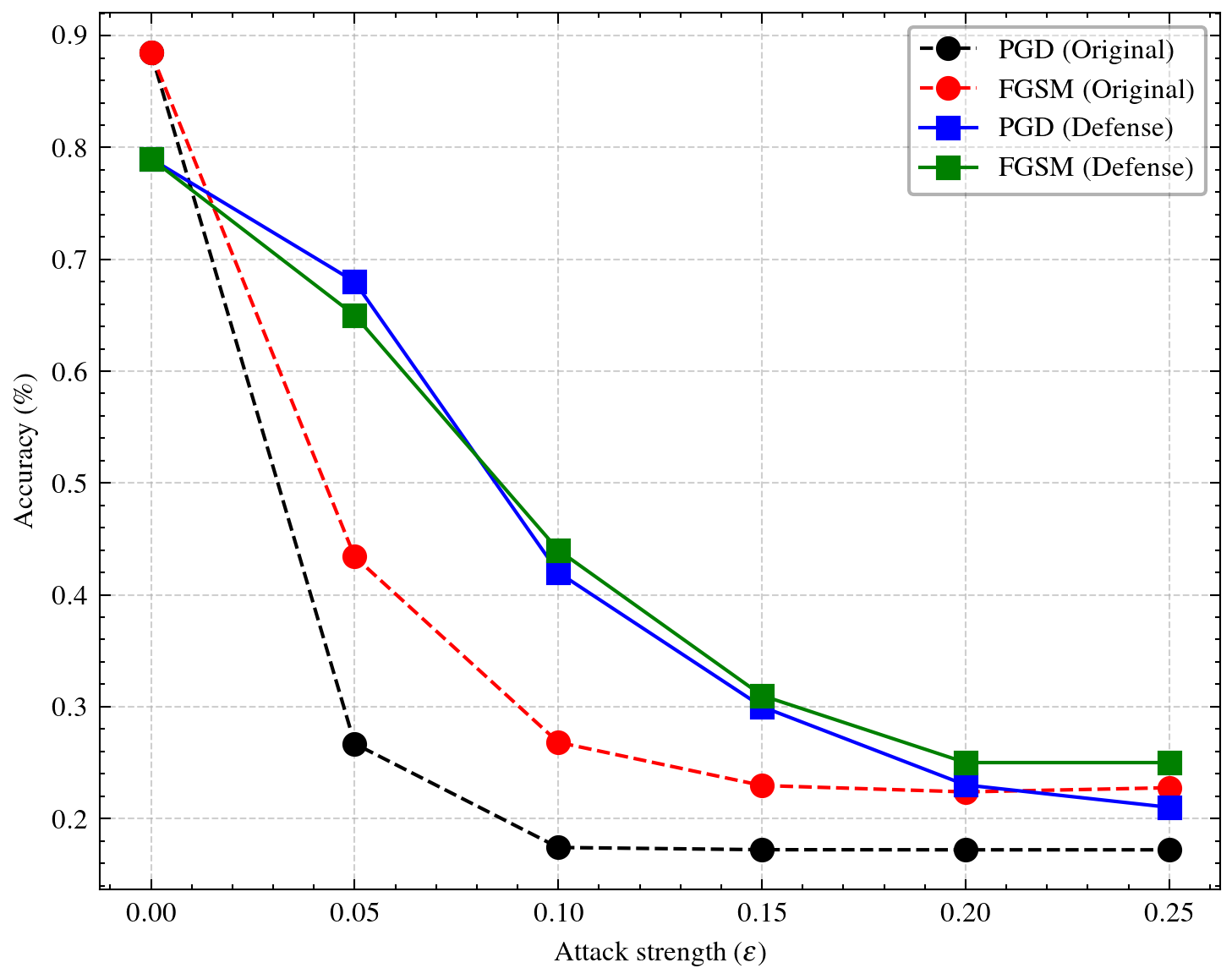}
        \caption{QNN-FashionMNIST (multi-qubit)}
        \label{fig:qnn_fashionmnist_full}
    \end{subfigure}
    \begin{subfigure}{0.32\linewidth}
        \centering
        \includegraphics[width=\linewidth]{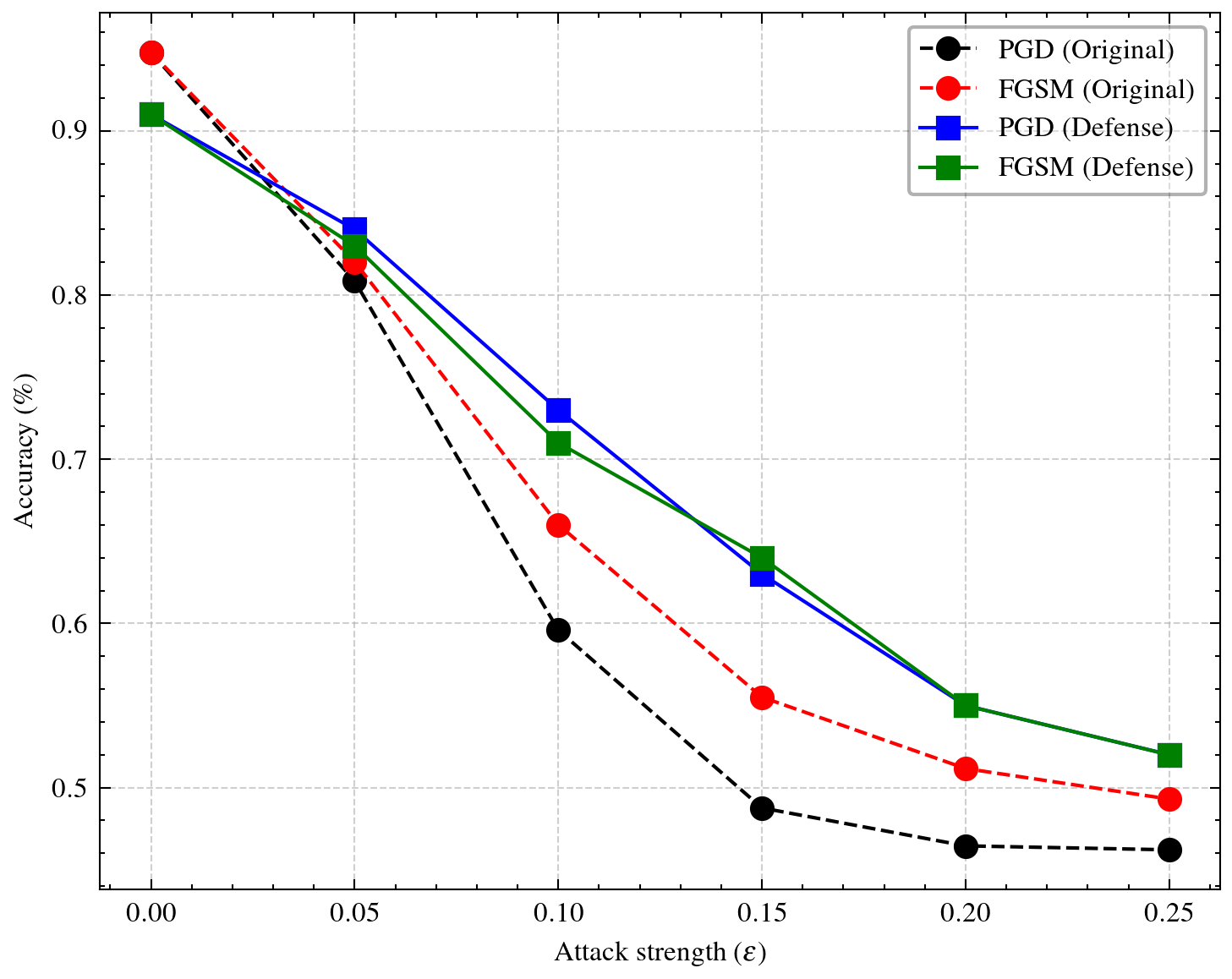}
        \caption{QNN-KMNIST (multi-qubit)}
        \label{fig:qnn_kmnist_full}
    \end{subfigure}
    
\vspace{0.2in}

    \begin{subfigure}{0.32\linewidth}
        \centering
        \includegraphics[width=\linewidth]{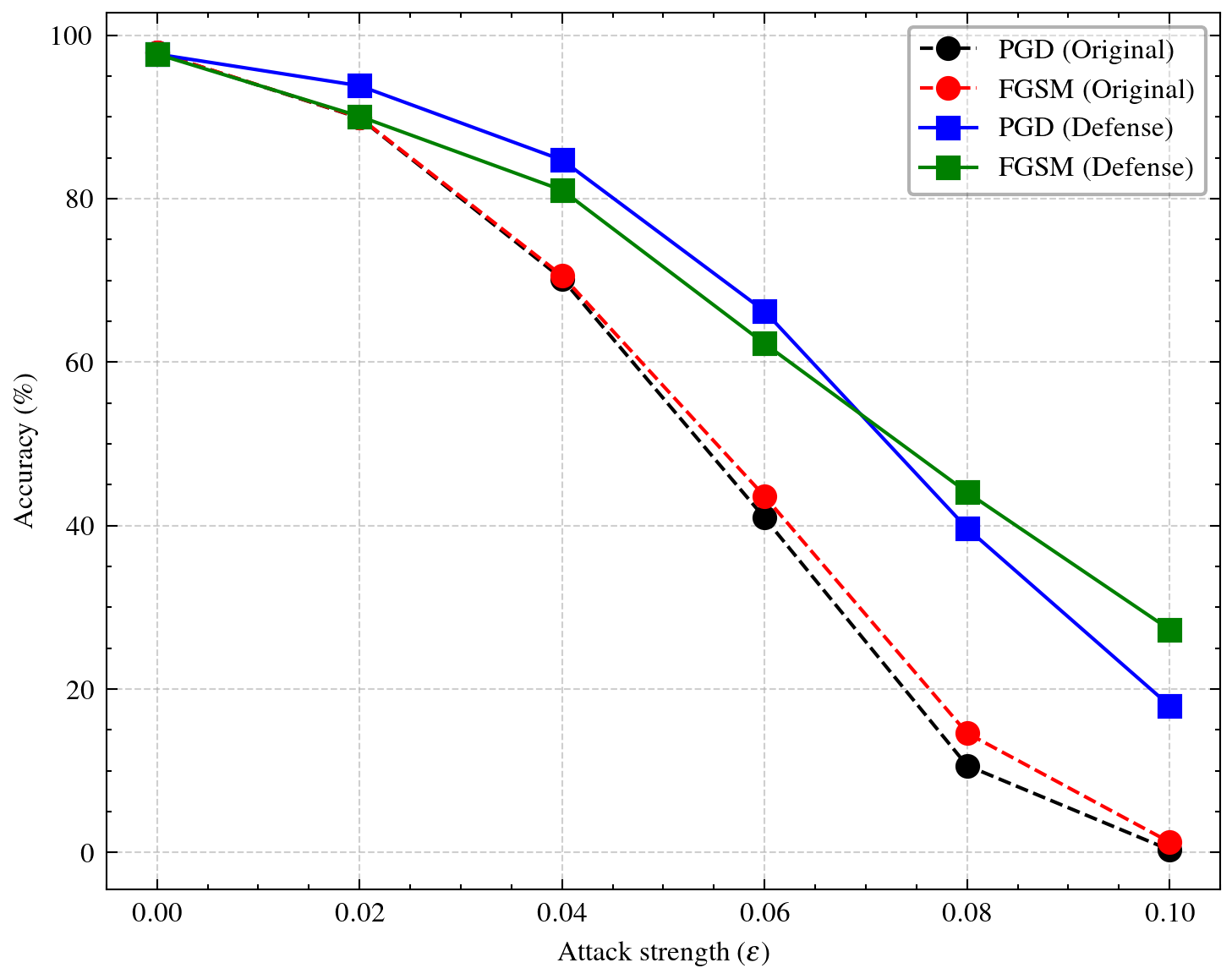}
        \caption{QCNN-MNIST}
        \label{fig:qcnn_mnist}
    \end{subfigure}
    \begin{subfigure}{0.32\linewidth}
        \centering
        \includegraphics[width=\linewidth]{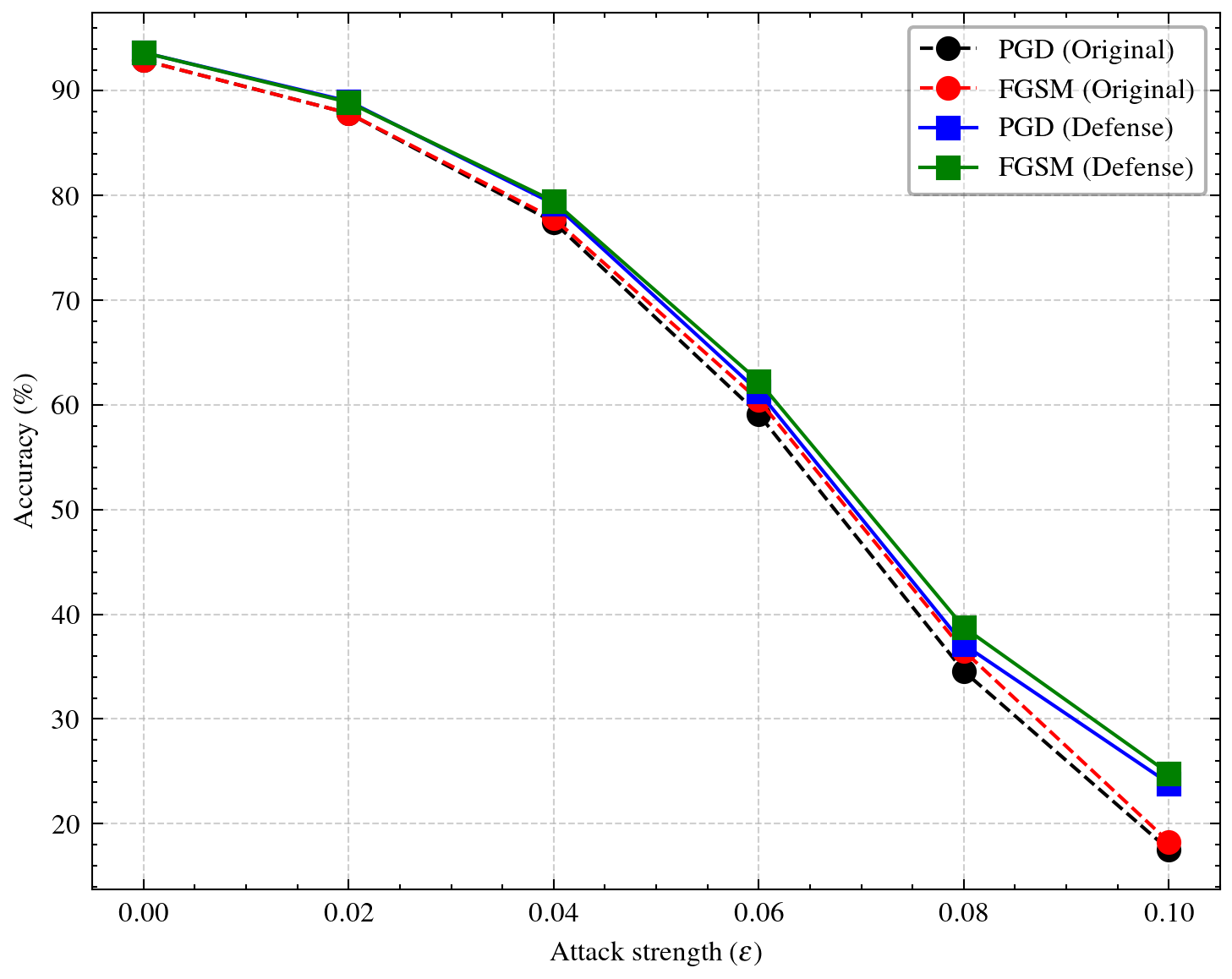}
        \caption{QCNN-FashionMNIST}
        \label{fig:qcnn_fashionmnist}
    \end{subfigure}
    \begin{subfigure}{0.32\linewidth}
        \centering
        \includegraphics[width=\linewidth]{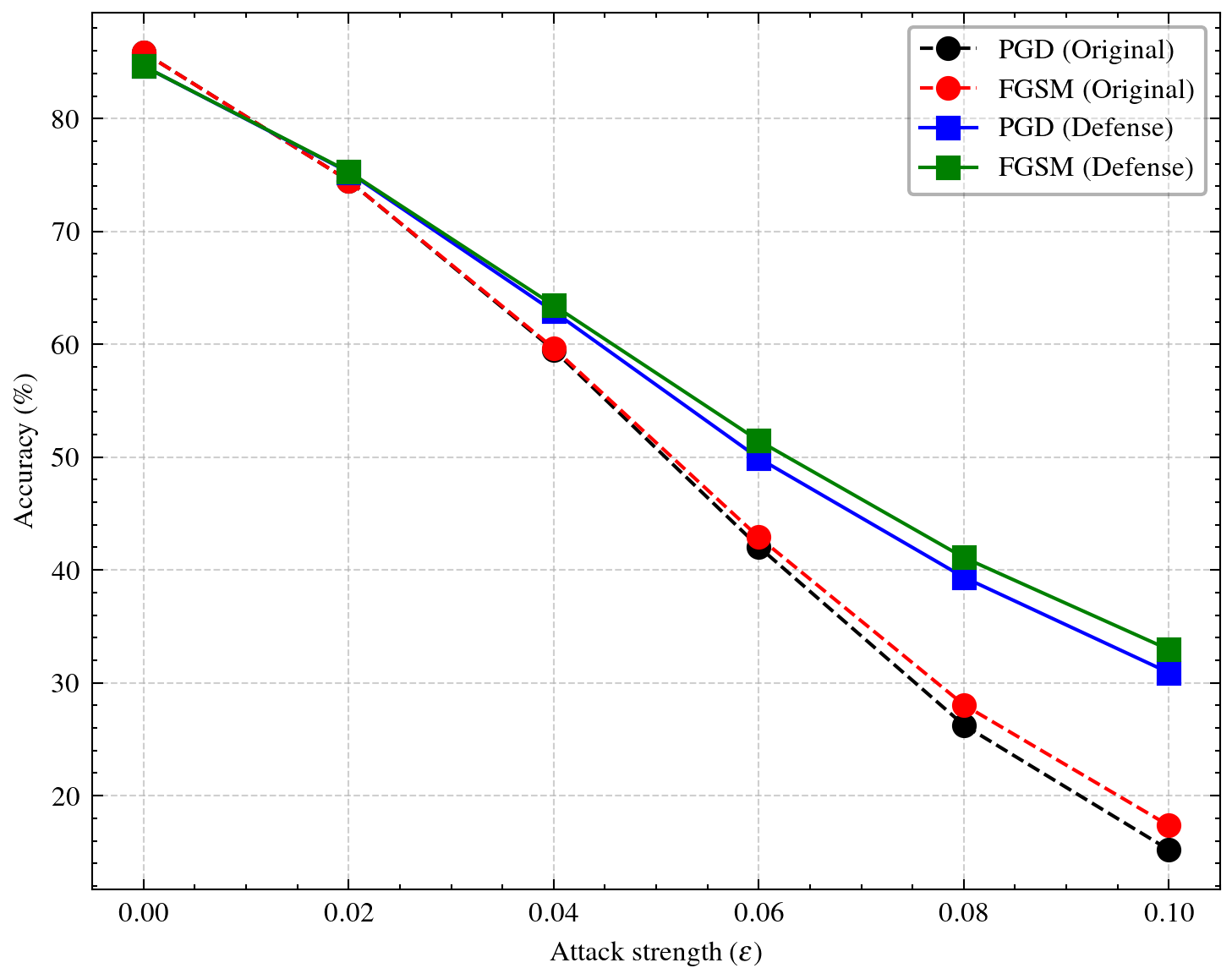}
        \caption{QCNN-KMNIST}
        \label{fig:qcnn_kmnist}
    \end{subfigure}
    
\vspace{0.2in}

    \begin{subfigure}{0.32\linewidth}
        \centering
        \includegraphics[width=\linewidth]{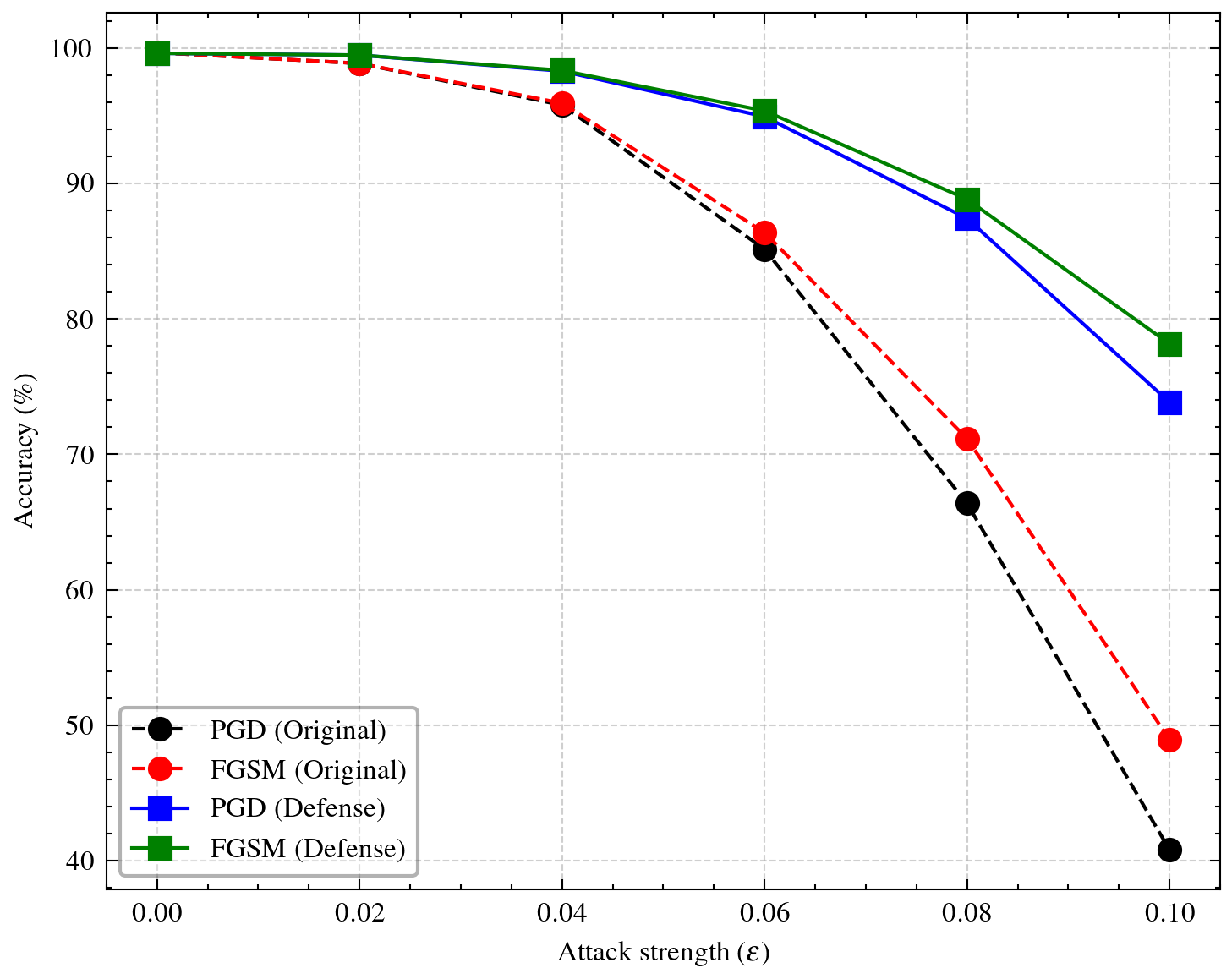}
        \caption{VQC-MNIST}
        \label{fig:vqc_mnist}
    \end{subfigure}
    \begin{subfigure}{0.32\linewidth}
        \centering
        \includegraphics[width=\linewidth]{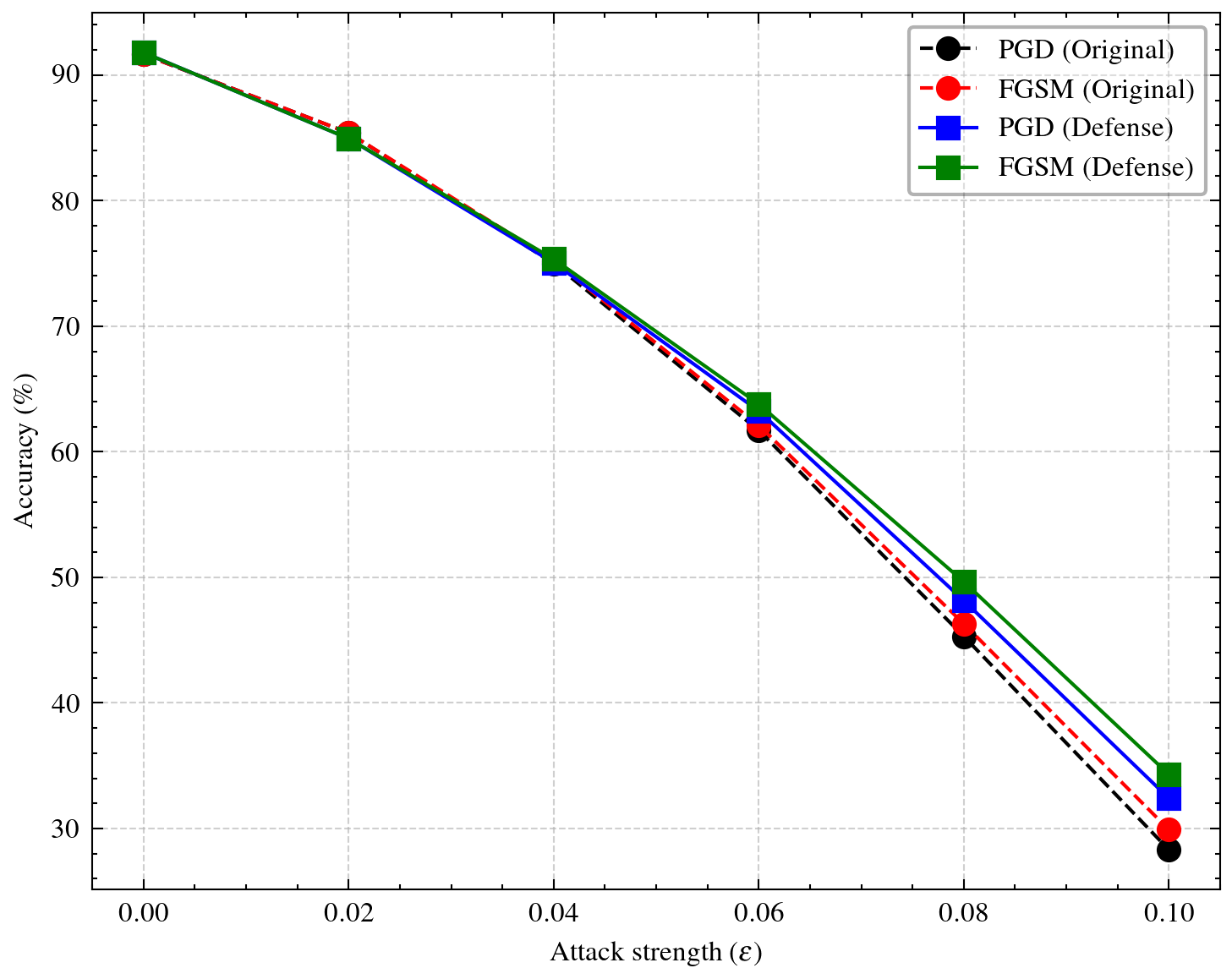}
        \caption{VQC-FashionMNIST}
        \label{fig:vqc_fashionmnist}
    \end{subfigure}
    \begin{subfigure}{0.32\linewidth}
        \centering
        \includegraphics[width=\linewidth]{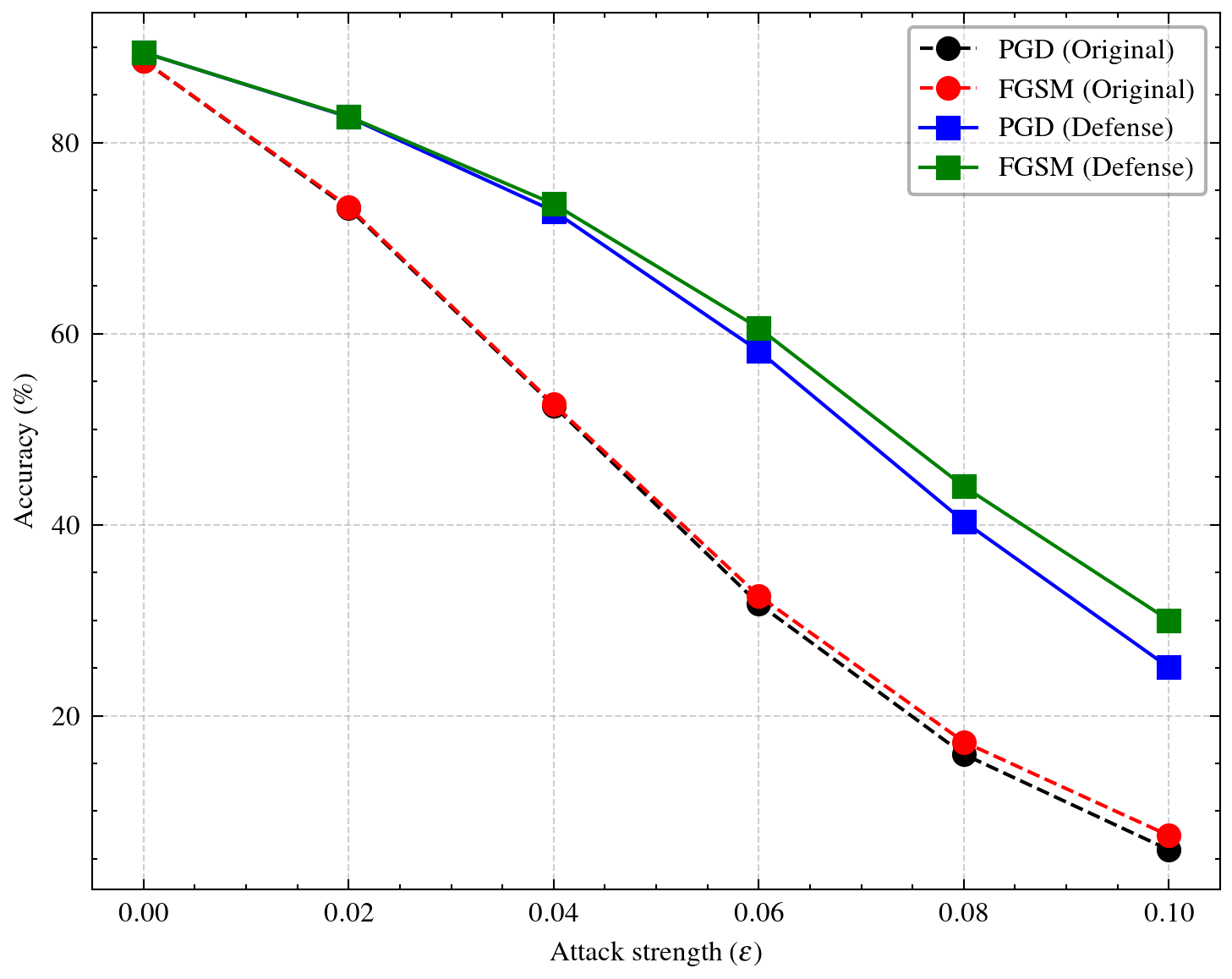}
        \caption{VQC-KMNIST}
        \label{fig:vqc_kmnist}
    \end{subfigure}

    \caption{Model test accuracies under FGSM and PGD attacks across different models and datasets for different perturbation strengths. Each figure includes the adversarial accuracy with and without the proposed defense method. Each row corresponds to a model, and each column corresponds to a dataset.}
    \label{fig:attack_accuracies_all}
\end{figure*}

\subsection{Results for Single-Qubit Steering in Hybrid QML Models}

To evaluate the effectiveness of the proposed controlled passive steering-based state preparation as a replacement for angle encoding for adversarial defense, we perform FGSM and PGD attacks on the QNN both with and without the defense. 
In the defended QNN, each individual \(R_y\) rotation used in angle encoding was replaced with a passive steering circuit. As shown in the first row in Figure~\ref{fig:attack_accuracies_all}, the accuracy of the undefended model decreases as the perturbation strength \(\epsilon\) increases, reaching its minimum at the highest tested perturbation strength, \(\epsilon = 0.25\). Under the FGSM attack with \(\epsilon = 0.25\), the model accuracy drops to 0.65 for MNIST, corresponding to an 34\% accuracy reduction from the original clean accuracy, to 0.23 for FashionMNIST, corresponding to a 74\% reduction, and to 0.49 for KMNIST, corresponding to an 48\% reduction. Similarly, under the PGD attack, the maximum accuracy degradation also occurs at \(\epsilon = 0.25\), with reductions of 52\%, 80\%, and 51\% for MNIST, FashionMNIST, and KMNIST, respectively.

After applying the single-qubit controlled passive steering-based state preparation defense, the clean accuracy changes only marginally, decreasing by 2\% for MNIST, 8\% for FashionMNIST, and 6\% for KMNIST. However, as illustrated in the first row in Figure~\ref{fig:attack_accuracies_all}, the adversarial accuracy under both FGSM and PGD attacks improves significantly after applying the defense. In the attack scenario with the highest tested perturbation strength, \(\epsilon = 0.25\), the adversarial accuracy increases by 13\% for MNIST, 62\% for FashionMNIST, and 19\% for KMNIST under the FGSM attack. Under the PGD attack, the corresponding adversarial accuracy improvements are 44\%, 96\%, and 26\% for MNIST, FashionMNIST, and KMNIST, respectively. 

Overall, these results demonstrate that replacing single-qubit angle encoding with controlled passive steering-based state preparation provides an effective and robust defense against adversarial attacks, while preserving nearly the same clean accuracy as the original model.

\subsection{Results for Multi-Qubit Steering in Hybrid QML Models}

In this approach, all four \(R_y\) rotation gates in the quantum kernel are replaced with a single passive steering circuit. After fixing \(J\) to \(\pi/16\), the appropriate value of \(N\) was identified by analyzing the clean accuracy of the QNN model as shown in Table~\ref{tab:jn_pairs}. In contrast to single-qubit steering, this method requires more iterations because the passive steering circuit acts on all four qubits simultaneously as explained in Section~\ref{sec:defense_angle}. However, unlike single-qubit steering, which requires one ancilla qubit for each steering circuit and thus a total of four ancilla qubits, this approach requires only one ancilla qubit. Aside from this difference in the required \(N\) value, the defense behaves as expected, as demonstrated in the second row of Figure~\ref{fig:attack_accuracies_all}.

\subsection{Results for Multi-Qubit Steering in Fully-Quantum Models}

In this experiment, we evaluate two main aspects: the effect of the proposed defense on fully quantum models and its applicability to amplitude encoding. QCNN and VQC are used as the target models, and both FGSM and PGD attacks are performed for evaluation. To replace amplitude encoding with controlled passive steering-based state preparation, we adopt the same strategy used in multi-qubit steering in QNN, where the entire amplitude-encoding circuit is replaced with a single steering circuit. We then fix \(J\) to \(\pi/10\) and determine the corresponding \(N\) values, as summarized in Table~\ref{tab:jn_pairs}.

FGSM and PGD attacks are subsequently applied to the binary classification models. We observe that, compared with the QNN, both QCNN and VQC are more sensitive to adversarial perturbations. Therefore, we set the maximum perturbation strength to 0.1. As shown in the third and fourth rows in Figure~\ref{fig:attack_accuracies_all}, the proposed defense effectively improves adversarial accuracy while maintaining clean accuracy with only minimal degradation. Among the three evaluated datasets, FashionMNIST exhibits the smallest improvement. 

Table~\ref{tab:result_sum} presents a summary of the PGD adversarial attack results for QNN, QCNN, and VQC with perturbation strength \(\epsilon = 0.1\). Based on these results, we observe that QNN exhibits the largest reduction in clean accuracy, with a maximum drop of \(9.05\%\). However, QNN also achieves the highest gain in adversarial accuracy, reaching up to \(40.19\%\). In contrast, both QCNN and VQC show smaller reductions in clean accuracy. Overall, these results clearly demonstrate that controlled passive steering-based state preparation, used to replace the encoding stage, improves adversarial accuracy across different QML models.

\begingroup
\begin{table}[htbp]
\centering
\caption{Summary of clean accuracy reduction (plain text) and adversarial accuracy improvement (bold text) under PGD attacks with $\epsilon = 0.1$. The clean accuracy reduction is calculated as the difference between the clean accuracy of the undefended model and that of the defended model, while the adversarial accuracy improvement is calculated as the difference between the adversarial accuracy of the defended and undefended models, across datasets and models. Here, QNN (S) denotes the single-qubit steering setting, and QNN (M) denotes the multi-qubit steering setting.}
\label{tab:result_sum}
\small
\begin{tabular}{lccc}
\toprule
\multirow{2}{*}{\textbf{Model}} & \multicolumn{3}{c}{\textbf{Datasets}} \\
\cmidrule(lr){2-4}
 & \textbf{MNIST} & \textbf{FashionMNIST} & \textbf{KMNIST} \\
\midrule
QNN (S)  & (-1.67, \textbf{6.50}) & (-7.84, \textbf{40.19}) & (-6.07, \textbf{16.96}) \\
QNN (M)  & (-0.89, \textbf{0.01}) & (-9.05, \textbf{24.59}) & (-3.77, \textbf{13.41}) \\
QCNN & (-0.19, \textbf{17.64}) & (0.75, \textbf{6.35}) & (-1.2, \textbf{15.7}) \\
VQC  & (-0.05, \textbf{33.05}) & (-0.05, \textbf{4.55}) & (0.85, \textbf{19.1})  \\
\bottomrule
\end{tabular}
\end{table}
\endgroup

\section{Conclusion} \label{sec:conclusion}

This paper presented a measurement-induced steering-based defense for quantum machine learning (QML), aimed at improving robustness against adversarial perturbations by replacing the conventional quantum encoding stage with a controlled passive steering-based state-preparation process. Across different QML models, datasets, and encoding schemes, the results show that appropriately tuning the steering strength and the number of steering iterations can substantially improve adversarial accuracy by suppressing adversarial noise while preserving clean-data performance, with adversarial accuracy improvements of up to 40.19\%. These findings highlight passive steering as a practical mechanism applicable to multiple representative hybrid and fully quantum classifiers for stabilizing quantum representations before inference, thereby offering a new direction for building more reliable and secure QML systems against adversarial attacks.

\balance

\bibliographystyle{IEEEtran}
\bibliography{refs.bib}

\end{document}